\documentclass[twocolumn,aps,pre,longbibliography]{revtex4-2}
\usepackage{epsfig}
\usepackage{natbib}
\usepackage{ulem}
\usepackage{xspace}
\usepackage{xcolor}
\usepackage{bbold}
\usepackage{float}
\usepackage{amsmath}
\usepackage{hyperref}
\hypersetup{
	colorlinks=true,
	linkcolor=blue,
	citecolor=blue,
	urlcolor=blue
}

\newcommand{\Ob}{\mathcal{O}}
\newcommand{\Z}{\mathcal{Z}_t}
\newcommand{\new}[1]{{\color{black}#1}}

\begin{document}

	\title{Brownian particles in periodic potentials: coarse-graining versus fine structure}
	
		\author{Lucianno Defaveri$^1$}
		\author{Eli Barkai$^{2}$}
		\author{David A. Kessler$^{1}$}
		
		\affiliation{$^1$Department of Physics, Bar-Ilan University, Ramat Gan 52900, Israel}
		\affiliation{$^2$Department of Physics, Institute of Nanotechnology and Advanced Materials, Bar Ilan University, Ramat Gan 52900, Israel}

	\begin{abstract}
		
We study the motion of an overdamped particle connected to a thermal heat bath in the presence of an external periodic potential in one dimension. 
When we coarse-grain, i.e., bin the particle positions using bin sizes that are larger than the periodicity of the potential, the packet of spreading particles, all starting from a common origin, converges to a normal distribution centered at the origin with a mean-squared displacement that grows as $2 D^* t$, with an effective diffusion constant that is smaller than that of a freely diffusing particle.
We examine the interplay between this coarse-grained description and the fine structure of the density, which is given by the Boltzmann-Gibbs (BG) factor $e^{-V(x)/k_B T}$, the latter being non-normalizable.
We explain this result and construct a theory of observables using the Fokker-Planck equation. 
These observables are classified as those that are related to the BG fine structure, like the energy or occupation times, while others, like the positional moments, for long times, converge to those of the large-scale description. Entropy falls into a special category as it has a coarse-grained and a fine structure description.
The basic thermodynamic formula $F=TS - E$ is extended to this far-from-equilibrium system.
The ergodic properties are also studied using tools from infinite ergodic theory.
		
	\end{abstract}

	\maketitle
	
	\section{Introduction}
	
\begin{figure}[H]
		\includegraphics[width = 0.48 \textwidth]{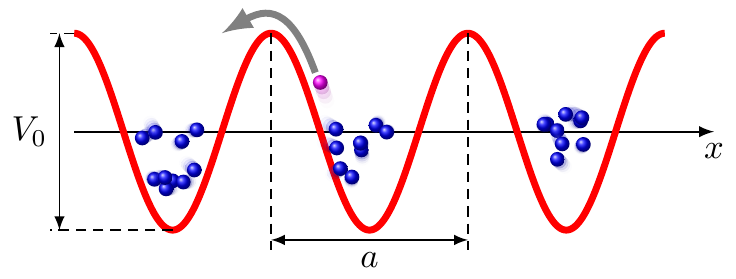}
		\caption{A schematic representation of the Brownian motion \new{of non-interacting particles,} using the potential in Eq.\,({\ref{eq:cosine-pot-def}}). The lattice period is $a$, the height of the potential barrier between wells is $V_0$ and the system is at temperature $T$. \label{fig:model}}
	\end{figure}

	Problems involving diffusion of atoms and molecules on surfaces, lattices, and general periodic potentials have been studied for decades \cite{Lifson1962,Reimann2001,Reimann2002,Ala-Nissila2002,Denisov2014,Dean2014,Lips2018,Radhakrishnan2022,Kim2022,Antonov2022,Antonov_2_2022} due to their applicability to a wide range of systems such as diffusion of adatoms \cite{Ehrlich1966,Ala-Nissila2002}, of proteins on a membrane \cite{Reister-Gottfried2010} and in one dimensional corrugated channels \cite{Reguera2001,Spiechowicz2016,Yang2017,Mangeat2018,Li2019,Li2020,Dagdug2021,Alexandre2022,Breoni2022}.
	Brownian particles in a one-dimensional periodic potential landscape $V(x)$, stretching across all space $(-\infty,\infty)$, cannot reach a state of equilibrium since, due to the nonbinding nature of the potential, the equilibrium distribution is not normalized as $\int_{-\infty}^{\infty}e^{-{V(x)}/{k_B T}} dx \to \infty$, where $k_B$ is the Boltzmann constant and $T$ the temperature of the environment. For short times, particles moving in a periodic lattice become stuck in attractive regions, or wells, of the potential. Eventually, however, the particles will experience an environmental fluctuation large enough to overcome the finite potential barrier and will reach a neighboring well \cite{Hanggi1990,Ferrando1993}. A schematic representation of this model is shown in Fig. \ref{fig:model}. This macroscopic motion is characterized by an effective diffusion constant $D^*$ which is always smaller than the free diffusion constant $D$  \cite{Lifson1962}. 
	
	Despite these types of systems being unable to reach a state of true equilibrium and therefore not obeying the ergodic hypothesis, the  Boltzmann-Gibbs factor, though non-normalizable, can still be used to study the properties of the system, as is the case with other non-confining potentials \cite{Aghion2019,Aghion2020,Farago2021,Farago2021}, logarithmic potentials used in subrecoil-laser-cooled gases \cite{Barkai2021}, diffusion processes with heterogeneous diffusion fields \cite{Leibovich2019,Wang2019} and random potentials used in Sinai diffusion \cite{Padash2022}. We show here that this non-normalizable state, i.e., the Boltzmann-Gibbs factor, gives the fine structure of the probability packet, and discuss the consequences of this. 
	This non-normalized state was foreseen by Sivan and Farago \cite{Sivan2018,Sivan2019}. By fine structure, we mean the density fluctuations on the scale of the period of the potential, which, in the long time limit, is of course much smaller than the scale associated with diffusion $\sqrt{2 D^* t}$.
	
	Experimental advances in optical lattices \cite{Lutz2013,Spar2022} which allow experimentalists to probe the fine-grained nature of systems, motivate us to ask: how does the interplay between the fine structure and more coarse-grained descriptions, which are both present in the probability density function (PDF), affect the properties of observables? What are their ergodic properties? Our aim in this paper is to answer those questions.
	It should be noted that one may observe the density of the spreading packet of particles either at a coarse-grained level or by paying attention to the fine structure. That it is say, when we observe the concentration of many non-interacting particles in the periodic potential, we may bin the data with bin sizes either smaller or greater than the period of the lattice. The latter case, which we call coarse-graining, will lead to the loss of information, though it is sometimes needed, since in the long time limit, within a small bin, we may not find a statistically sufficient number of particles. The coarse-graining issue is then translated to other observables, like entropy. As explained below, it can generally result in widely different points of view on the system if compared to a fine-scale observation.
	
	 We note that considerable attention was devoted in the literature to the coarse-graining problem, in a thermodynamical setting \cite{Esposito2012,Alonso-Serrano2017,Dechant2019,Busiello2019,Martinez2019,Chakraborti2022,Dieball2022}, here, however, we deal with a new domain, that of infinite ergodic theory \cite{Aghion2019,Aghion2020,Farago2021,Aaronson2005,Thaler2006,Akimoto2013,Giordano2022}. As we explain below, the \new{time-invariant infinite} density in our system is the Boltzmann-Gibbs factor, which, as we mentioned, is non-normalizable.
	
	The manuscript is organized as follows. In Section \ref{sec:model} we describe the potential and the basic concepts and tools of our model. In Section \ref{sec:asymptotic-solution} we present, using intuitive arguments, the long-time PDF of the Brownian particle. We discuss the different types of observables with respect to their ensemble averages in Section \ref{sec:observables}, and with respect to their time, together with their ergodic properties, in Section \ref{sec:ergodicity}. In Section \ref{sec:thermodynamics} we calculate the entropy for both coarse-grained and fine structure descriptions. In Section \ref{sec:eigen} we provide a rigorous derivation of the PDF using an eigenfunction expansion. Finally, in Section \ref{sec:final-remarks} we present our concluding remarks.
	
	\section{Model \label{sec:model}}

	We consider the one-dimensional overdamped motion of a Brownian particle in a thermal environment of temperature $T$ which is also subjected to the external periodic potential $V(x) = V(x+a)$, consisting of attractive well regions (local minima) separated by potential barriers (local maxima) of height $V_0$. A potential that fulfills these characteristics is given by
	\begin{eqnarray}
		V(x) = - \frac{V_0}{2} \cos(2 \pi x/a) \label{eq:cosine-pot-def} \, ,
	\end{eqnarray}
	where $a$ is the lattice spacing. In Fig.\,\ref{fig:model} we show a schematic representation of the model. The probability density $P_t(x)$ of the particle at time $t$ is described by the Fokker-Planck equation (FPE) \cite{Risken1989}
	\begin{eqnarray}
		\frac{\partial P_t(x)}{\partial t} &=& D \left\{ \frac{\partial^2 P_t(x)}{\partial x^2} + \frac{1}{k_B T} \frac{\partial}{\partial x} \left[ \frac{\partial V}{\partial x} P_t(x) \right] \right\} \, , \label{eq:FPE}
	\end{eqnarray}
	where $D$ is the bare diffusion constant and $T$ is the temperature of the environment.

	Equivalently, we could describe the system at the level of individual trajectories, or realizations, using the Langevin equation
	\begin{eqnarray}
		\gamma \dot x &=& -  \frac{\partial V}{\partial x} + \sqrt{2 \gamma k_B T} \, \eta(t) \, , \label{eq:langevin}
	\end{eqnarray}
	where $\gamma$ is the damping constant, which obeys Einstein's relation $D = k_B T/\gamma$, with $\eta(t)$ being a stochastic Gaussian white noise with zero mean and variance $\langle \eta(t) \eta(t') \rangle = \delta(t-t')$. For each realization, we would have a stochastic trajectory $x_\eta$, so that
	\begin{eqnarray}
		P_t(x) = \langle \delta(x - x_\eta) \rangle_\eta \, ,
	\end{eqnarray}
	where the brackets $\langle ... \rangle_\eta$ represent averages taken over an ensemble of trajectories $x_\eta$. We will later use the Langevin equation to numerically compute the time averages of physical observables, while in the first part of the manuscript we will use the Fokker-Planck equation.

    An important dimensionless control parameter for studying the system is the ratio between the height of the potential barrier and the typical energy from thermal fluctuations $V_0/k_B T$. 
	Our main results are valid in all temperature ranges, provided that the time is large enough.
	
	\section{Asymptotic solution \label{sec:asymptotic-solution}}
	
	In Section\,\ref{sec:eigen} we present a derivation, using an eigenfunction expansion, of the asymptotic solution for the PDF $P_t(x)$ governed by Eq.\,(\ref{eq:FPE}).
	For the moment, we will rely on the more physically transparent ansatz-based derivation of Sivan and Farago~\cite{Sivan2018,Sivan2019} which we herein recapitulate in order to make the current work self-contained.   For long times, the mean squared displacement, which is equivalent to the second positional moment, follows the expression $\langle x^2 \rangle \sim 2 D^* t $, where $D^*$ is the effective diffusion constant, which can be calculated, as shown by Lifson and Jackson \cite{Lifson1962}, as
	\begin{eqnarray}
		D^* = \frac{D}{\left\langle e^{\frac{V(x)}{k_B T}} \right\rangle_a \left\langle e^{-\frac{V(x)}{k_B T}} \right\rangle_a}  \, , \label{eq:liffson}
	\end{eqnarray}
	where we define the average over a lattice period as $\langle f \rangle_a = (1/a) \int_{-a/2}^{a/2} f(x) dx$. For the specific case of the potential in Eq.\,(\ref{eq:cosine-pot-def}) we have $D^* = D/{I_0^2\left({V_0}/{2 k_B T}\right)}$, with $I_0(...)$ being the 0-th modified Bessel function of the first kind. This allows us to define the effective diffusive lengthscale $\sqrt{2 D^* t}$.
	
	 Any periodic potential $\bar{V}(x)$ can be shifted by a constant value $\delta V \equiv ({k_B T}/{2}) \ln \left[ \left\langle e^{{-V}/{k_B T}} \right\rangle_a / \left\langle e^{{V}/{k_B T}} \right\rangle_a \right]$, giving a new potential $V(x) = \bar{V}(x) + \delta V$. For this new potential, we have that $\left\langle e^{{V}/{k_B T}} \right\rangle_a = \left\langle e^{-{V}/{k_B T}} \right\rangle_a$. For simplicity, we will use this convention and study potentials that obey this equality, as the force field is clearly invariant under the above-mentioned transformation, and therefore Eqs.\,(\ref{eq:FPE}) and (\ref{eq:langevin}) are unchanged.
	
\begin{figure}
		\includegraphics[width = 0.475 \textwidth]{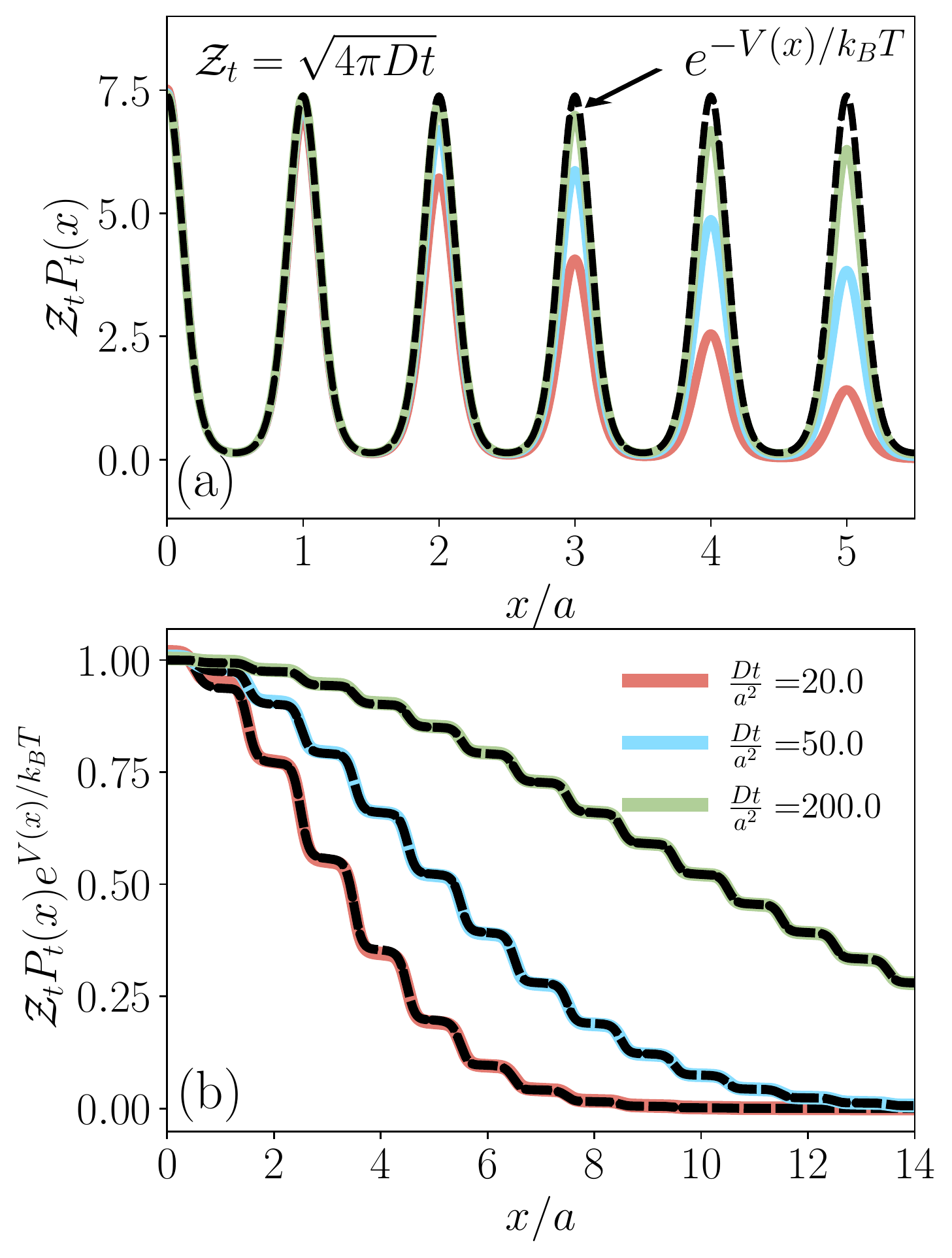}
		\caption{\new{Panel (a): numerical results for $\mathcal{Z}_t P_t(x)$ (solid lines), for the three different times shown in the legend of panel (b). The black dashed line represents the Boltzmann-Gibbs factor $e^{- V(x)/k_B T}$. We can observe that, for longer times and for $x \ll \sqrt{4 D^* t}$, $\mathcal{Z}_t P_t \approx e^{- V(x)/k_B T}$, as expected from Eq.\,(\ref{eq:infinite-density-def}). Panel (b): numerical results for $\mathcal{Z}_t P_t(x)$ divided by the Boltzmann-Gibbs factor (solid lines) for three different times shown in the legend. The black dashed lines represent the prediction in Eq. (\ref{eq:Pt-theoretical}). In both panels $V_0/k_B T = 4$  and $D^*/D \approx 0.19$.}
		\label{fig:model-infiniteP}}
	\end{figure}

	For long times and a range of positions much less than the diffusive lengthscale, $x \ll \sqrt{2 D^* t}$, the PDF becomes proportional to the BF distribution as
	\begin{eqnarray}
	    P_t(x) \propto  \, \frac{e^{- \frac{V(x)}{k_B T}} }{t^{\alpha}} \label{eq:fine-structure-PDF} \, .
	\end{eqnarray}
	where $\alpha > 0$.  We see that this is a solution by plugging Eq.\,(\ref{eq:fine-structure-PDF}) in the FPE (\ref{eq:FPE}), the right-hand side is identically zero and the left-hand side is $\partial_t P_t(x) \approx \alpha P_t(x)/t$, and in the limit of large $t$ we have $\partial_t P_t(x) \to 0$. For large lengthscales, $x \sim \sqrt{2 D^* t}$, the fine structure of the PDF can be neglected, leading to a free particle-like description, with an effective diffusion constant $D^*$, that is,
	\begin{eqnarray}
	    P_t(x) \approx \frac{e^{- \frac{x^2}{4 D^* t}}}{\sqrt{4 \pi D^* t}} \, . \label{eq:coarsed-PDF}
	\end{eqnarray}
	We compare Eq. (\ref{eq:fine-structure-PDF}) and Eq. (\ref{eq:coarsed-PDF}), to conclude that $\alpha = 1/2$. By matching both limits we obtain a uniform approximation as
	\begin{eqnarray}
	    P_t(x) \approx \mathrm{const} \, \frac{e^{- \frac{V(x)}{k_B T}} e^{- \frac{x^2}{4 D^* t}}}{\sqrt{4 \pi D^* t}} \, .
	\end{eqnarray}
	The constant is calculated by imposing the normalization of the PDF,
	\begin{eqnarray}
	     \mathrm{const}  \int_{-\infty}^\infty \frac{e^{- \frac{V(x)}{k_B T}} e^{- \frac{x^2}{4 D^* t}}}{\sqrt{4 \pi D^* t}} dx \approx 1 \, .
	\end{eqnarray}
	We perform a change of variables to $y \equiv x / \sqrt{t}$, 
	\begin{eqnarray}
	     \mathrm{const}  \int_{-\infty}^\infty \frac{e^{- \frac{V(y \sqrt{t})}{k_B T}} e^{- \frac{y^2}{4 D^*}}}{\sqrt{4 \pi D^*}} dy \approx 1 \, ,
	\end{eqnarray}
	where we see that the Boltzmann-Gibbs factor $e^{- {V(y \sqrt{t})}/{k_B T}}$ oscillates rapidly allowing it to be replaced by its average value in a period, that is,
	\begin{eqnarray}
	     \mathrm{const} \left\langle e^{- \frac{V}{k_B T}} \right\rangle_a  \int_{-\infty}^\infty \frac{ e^{- \frac{y^2}{4 D^*}}}{\sqrt{4 \pi D^*}} dy \approx 1 \, , 
	\end{eqnarray}	
	where the integral is clearly unity, and $\mathrm{const} = {1}/{\left\langle e^{- {V}/{k_B T}} \right\rangle_a}$. The uniform approximation becomes
	\begin{eqnarray}
	    P_t(x) & \approx & \frac{e^{- \frac{V(x)}{k_B T}} e^{- \frac{x^2}{4 D^* t}}}{\Z} \, , \label{eq:Pt-theoretical-0}
	\end{eqnarray}
	where we define the normalizing term
	\begin{eqnarray}
	    \Z \equiv \left\langle e^{- {V}/{k_B T}} \right\rangle_a \sqrt{4 \pi D^* t} = \sqrt{4 \pi D t} \, . \label{eq:Zt}
	\end{eqnarray}
	Using this uniform approximation, it is possible to obtain an \new{time-invariant infinite} density of the system as
	\begin{eqnarray}
		\lim_{t \to \infty} {\cal Z}_t P_t(x) =  e^{-\frac{V(x)}{k_B T}} \label{eq:infinite-density-def} \, ,
	\end{eqnarray}
	a result that is known for asymptotically flat potentials \cite{Aghion2019,Aghion2020}, which is here seen to also be valid in the case of periodic potentials. For finite long times, Eq.\,(\ref{eq:infinite-density-def}) holds for $x \ll \sqrt{2 D^* t}$, that is, $x$ much smaller than the diffusive lengthscale. This expression, which is valid regardless of initial conditions, shows that the system relaxes to a state closely related to thermal equilibrium described by the Boltzmann-Gibbs factor, even if the latter is non-normalized, with the time-dependent $\Z$ defined in Eq.\,(\ref{eq:Zt}) replacing the usual normalizing partition function. In panel (a) of Fig\,\ref{fig:model-infiniteP} we show the relaxation of $P_t(x)$ to the Boltzmann-Gibbs factor using a numerical integration of the FPE\,(\ref{eq:FPE}).

	The uniform approximation can be improved by considering additional long-time corrections. In Section \ref{sec:eigen} we present a rigorous eigenfunction derivation, while in this section we will follow the same principle used by Sivan and Farago in \cite{Sivan2018,Sivan2019} and propose a solution in the form
	\begin{eqnarray}
	    P_t(x) &=& \frac{e^{- \frac{V(x)}{k_B T} - \frac{x^2}{4 D^* t}}}{\Z} \left(1 - \frac{\tau(x)}{2 t} \right) \, , \label{eq:ansatz}
	\end{eqnarray}
	where $\tau(x)$ is an ansatz. We plug the proposed solution in Eq. (\ref{eq:ansatz}) into the FPE (\ref{eq:FPE}), and limit ourselves to long time contributions up to $O(t^{-3/2})$. The left-hand side of the FPE, in this limit, becomes,
	\begin{eqnarray}
	    \frac{\partial P_t(x)}{\partial t} &\approx& - \frac{e^{- \frac{V(x)}{k_B T}}}{2t \, \Z} \, ,
	\end{eqnarray}
	and the right-hand side of the FPE,
	\begin{eqnarray}
	    D \left( \frac{\partial^2 P_t(x)}{\partial x^2} + \frac{V'(x)}{k_B T} \frac{\partial P_t(x)}{\partial x}\right) =  - \frac{D}{D^*} \frac{e^{- \frac{V(x)}{k_B T}}}{2t \, \Z} \left( D^* \tau''(x) \right. \nonumber \\
	    + 1  \left. - \frac{V'(x)}{k_B T} \left[ x + D^* \tau'(x) \right]  \right) \, , ~~~~~~~~
	\end{eqnarray}
	leading to a differential equation for the ansatz as
    \begin{eqnarray}
         \tau''(x) - \frac{1}{D^*}\frac{V'(x)}{k_B T} \big(x + D^* \tau'(x) \big) = \frac{1}{D} - \frac{1}{D^*} \, .
    \end{eqnarray}
	This equation can be solved as
	\begin{eqnarray}
	    \tau(x) &=& \frac{1}{D} \int_0^x e^{\frac{V(y_1)}{k_B T}} \int_0^{y_1} e^{-\frac{V(y_2)}{k_B T}} dy_2 dy_1 + \nonumber \\
	    &~& - \frac{x^2}{2D^*} + \frac{C_0}{D} \int_0^x e^{\frac{V(y)}{k_B T}} dy + C_1 \, ,
	\end{eqnarray}
	where $C_1$ is a constant that  ensures the normalization of $P_t(x)$ and $C_0$ ensures that there is no biased particle flow.	For a symmetric potential and initial condition at the potential minimal, we expect the PDF to be distributed in space symmetrically, therefore we must have that $\tau(x) = \tau(-x)$, which leads to $C_0 = 0$. For an asymmetric potential, we must instead impose that there is no macroscopic drift of particles, that is, $C_0$ is defined to ensure that $\tau(a) = \tau(-a)$,
	\begin{eqnarray}
	    C_0 &=& \frac{1}{2 \int_{0}^a e^{\frac{V(y)}{k_B T}}dy} \left\{ \int_0^a e^{\frac{V(y_1)}{k_B T}} \int_{y_1}^a e^{-\frac{V(y_2)}{k_B T}} dy_2 dy_1  \right. \nonumber \\
	    &~& - \left. \int_0^a e^{\frac{V(y_1)}{k_B T}} \int_0^{y_1} e^{-\frac{V(y_2)}{k_B T}} dy_2 dy_1  \right\} \, . \label{eq:C0}
	\end{eqnarray}
	
	We can manipulate the expression for $\tau(x)$ to write that
	\begin{eqnarray}
	    \tau(x) & = &  \frac{x \, U_1(x)}{D^*} + \frac{U_2(x)}{D^*} \, , \label{eq:def-tau}
 	\end{eqnarray}
 	where $U_1(x)$ and $U_2(x)$ are $a$-periodic functions with 
 	\begin{eqnarray}
 	    U_1(x) &=& a  \frac{\int_0^x e^{\frac{V(y)}{k_B T}} dy}{\int_0^a e^{\frac{V(y)}{k_B T}} dy} - x \label{eq:U1-def}\, .
 	\end{eqnarray}
 	The initial conditions are present in $U_2(x)$, which we will here omit giving the full expression. The scaling $y \equiv x/\sqrt{t}$ represents the diffusive motion of the particles, it reflects the Gaussian spreading of the PDF, and we use this scaling to write
 	\begin{eqnarray}
 	    \frac{\tau(x)}{2t} = \frac{y U_1(x)}{2 D^* \sqrt{t}} + \frac{U_2(x)}{2t}    \, , 
 	\end{eqnarray} 
 	In this scale, we neglect terms of order $O(t^{-1})$, leaving us with the $y U_1(x)$ term, which contains contributions to both coarse-grained and fine-grained structures. We reach the final expression \cite{Sivan2018,Sivan2019}
    \begin{eqnarray}
        P_t(x) & \approx & \frac{e^{-\frac{V(x)}{k_B T}} e^{- \frac{x^2}{4 D^* t}}}{\mathcal{Z}_t} \left[ 1 - \frac{x \, U_1(x) }{2 D^* t} \right] \label{eq:Pt-theoretical} \, ,
    \end{eqnarray}
    where $\mathcal{Z}_t$, which was defined before, plays a similar role as the partition function for regular Boltzmann-Gibbs equilibrium.
    This final expression is valid regardless of the symmetry properties of $V(x)$ and can be used for any initial condition $x_0$ simply by translating the $x$-axis so that $x_0$ becomes the new origin.
    
	In Fig. (\ref{fig:model-infiniteP}) we compare our results in Eqs.\,(\ref{eq:infinite-density-def}) and (\ref{eq:Pt-theoretical}) with the numerical integration of Eq. (\ref{eq:FPE}). In the top panel (a), we show how the PDF $P_t(x)$ multiplied by $\Z$ converges in the long time limit to the Boltzmann-Gibbs factor, akin to systems with perfectly normalized BG states. In the lower panel, we plot the density divided by the Boltzmann-Gibbs factor versus $x$. In the long time limit, we expect a Gaussian propagator, similar to that of a free particle, with an effective diffusion constant $D^*$, however, at not too long times, the correction term in Eq.\,(\ref{eq:Pt-theoretical}) is clearly important.

	\section{Ensemble averages \label{sec:observables}}
	
	In this section, we focus on the ensemble average of a physical observable $\mathcal{O}(x)$ at a given time $t$, which we label $\langle \mathcal{O} \rangle_t$, given by
	\begin{eqnarray}
		\langle {\cal O} \rangle_t = \int_{-\infty}^{\infty} {\cal O}(x) P_t(x) dx \, .
	\end{eqnarray}
	We will now classify the different observables and their dependence on the non-normalized Boltzmann-Gibbs state in the long time limit. We will see that some observables are sensitive to the fine scale of the solution, namely, to the Boltzmann-Gibbs factor, while others are controlled by the coarse-grained description of $P_t(x)$, which amounts to a Gaussian.

    \subsection{Positional moments}
    
\begin{figure}
	\includegraphics[width = 0.47 \textwidth]{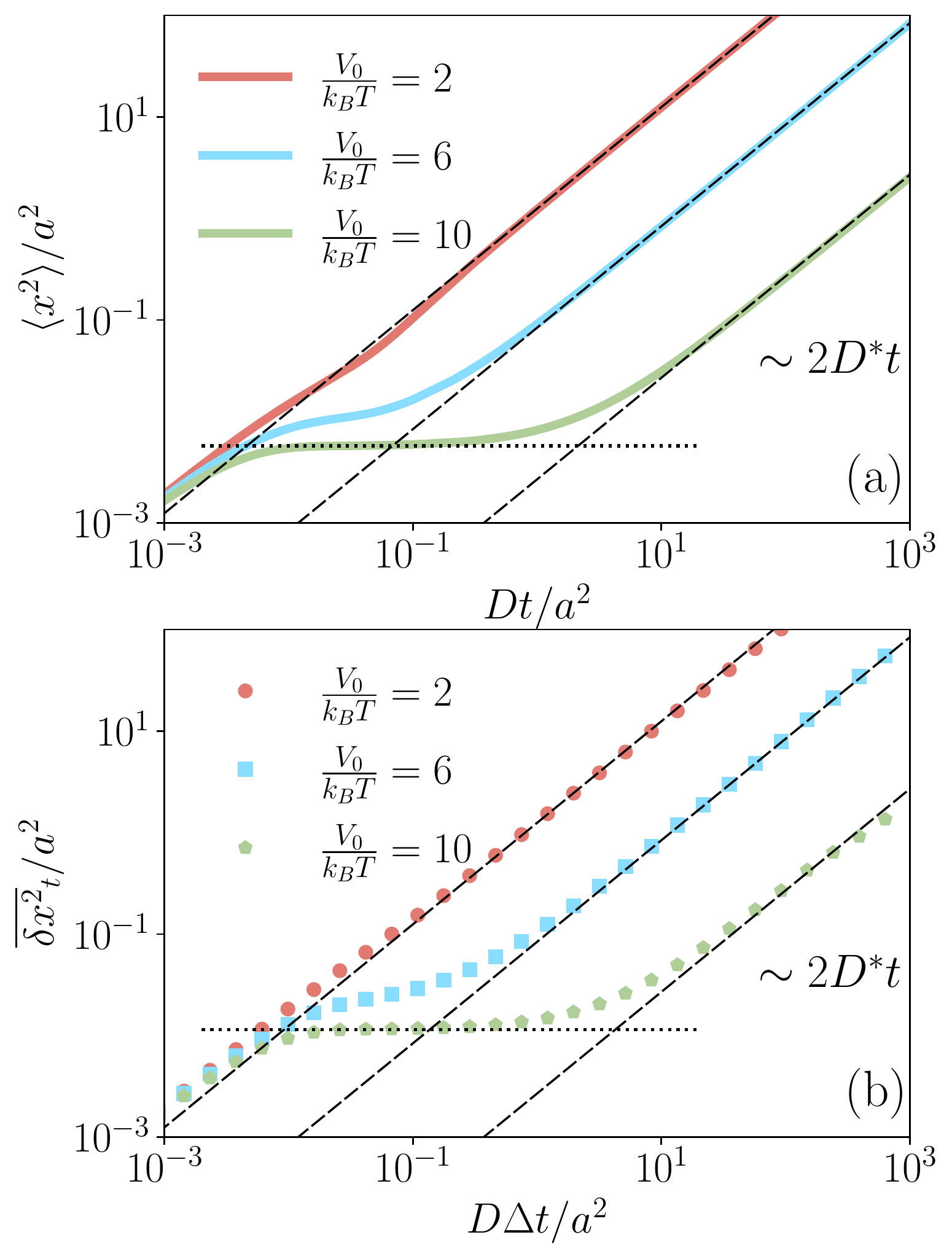}
	\caption{
	\new{Panel (a): the ensemble-averaged MSD (solid lines), for the different values of $V_0/k_BT$ shown in the legend, for particles starting at the origin.  Panel (b): the time-averaged MSD over a time interval $\Delta t$, as a function of the averaging window size, $t$, from a single trajectory (symbols), for the different values of $V_0/k_BT$ shown in the legend.
    The dashed lines correspond to $2 D^* t$, panel (a), Eq.\  (\ref{eq:msd}) and $2 D^*\Delta t$, panel (b), Eq.\  (\ref{eq:msd-long-time}), with the effective diffusion constant $D^*$, as given in Eq.\,(\ref{eq:liffson}).
    It is noticeable that, for the larger values of $V_0/k_BT$, where there is a range of times that are long but still shorter than the typical escape time, the system attains a quasi-stationary state, as given in Eqs.\,(\ref{eq:msd-nqe-ensemble}) and (\ref{eq:msd-nqe}), horizontal dotted lines in panels (a) and (b) respectively, before the long-time effective diffusion becomes dominant.}}
	\label{fig:moments}
	\end{figure}

    It is possible to calculate the $q$-th moments of $x$, $\langle |x|^q \rangle_t$, using the PDF in Eq.\,(\ref{eq:Pt-theoretical-0}). The statistical properties of these observables are controlled by the large-scale solution of the packet. For long times, their statistics follow those of a free particle with the effective diffusion constant $D^*$.
    As an example, we calculate the \new{ensemble average of the} second moment, the mean square displacement (MSD), $\langle x^2 \rangle_t$. We perform the same change variables to $y \equiv x/\sqrt{t}$ as we did to calculate the normalization in Section \ref{sec:asymptotic-solution}, to obtain the expression
    \begin{eqnarray}
        \langle x^2 \rangle_t & \approx & t \int_{-\infty}^{\infty} y^2 e^{- \frac{V(y \sqrt{t})}{k_B T}} \frac{e^{- \frac{y^2}{4 D^*}}}{\sqrt{4 \pi D}} dy \, .
    \end{eqnarray}
    In the long-time limit, we see that $e^{- {V(y \sqrt{t})}/{k_B T}}$ will oscillate rapidly, which allows us to replace its value for an average in a period, that is,
    \begin{eqnarray}
        \langle x^2 \rangle_t & \approx & t \, \left\langle e^{- \frac{V(x)}{k_B T}} \right\rangle_a \int_{-\infty}^{\infty} y^2 \frac{e^{- \frac{y^2}{4 D^*}}}{\sqrt{4 \pi D}} dy \nonumber \\
        & \approx & t \sqrt{\frac{D}{D^*}} \int_{-\infty}^{\infty} y^2 \frac{e^{- \frac{y^2}{4 D^*}}}{\sqrt{4 \pi D}} dy = 2 D^* t \label{eq:msd} \, ,
    \end{eqnarray}
    where we have used that $\langle e^{- {V}/{k_B T}} \rangle_a = \sqrt{{D}/{D^*}}$ and obtained the expected variance for normal diffusion with the effective diffusion constant $D^*$. This result can be extended to a general $q$-th moment as
    \begin{eqnarray}
        \langle |x|^q \rangle_t &\approx& \frac{2^q}{\sqrt{\pi}} \left( D^* t \right)^{\frac{q}{2}} \Gamma \left( \frac{1+q}{2} \right) \, , \label{eq:q-th-mom}
    \end{eqnarray}
    where $\Gamma(x)$ is the Gamma function. In Fig. \ref{fig:moments}(a) we plot the numerical evaluation of the mean square displacement, compared with their theoretical prediction in Eq.\,(\ref{eq:msd}). We have also observed numerically the validity of Eq.\,(\ref{eq:q-th-mom}) for long times (not shown).
    
    We may call observables like $|x(t)|^q$ coarse-grained observables since they are not sensitive to the fine structure, namely, the Boltzmann-Gibbs factor. In fact, Eq.\,(\ref{eq:q-th-mom}) are the moments of a perfectly normal Gaussian packet, with a variance given by Eq.\,(\ref{eq:msd}). Thus, as a standalone, in the long time limit, the moments in Eq.\,(\ref{eq:q-th-mom}) behave as those of a free particle with diffusion constant $D^*$, as mentioned.
    
    \new{For larger values of $V_0/k_B T$ and short times (shorter than the typical escape time), as we can see in Fig.\,\ref{fig:moments}(a), the particle will reach a transient quasi-stationary state \cite{Defaveri2020,Anteneodo2021}. The ensemble average will be equivalent to that of a particle in thermal equilibrium within a single site, that is,
    \begin{eqnarray}
        \langle x^2 \rangle_t \approx \frac{ \int_{-a/2}^{a/2} x^2 e^{-\frac{V(x)}{k_B T}} dx }{\int_{-a/2}^{a/2} e^{-\frac{V(x)}{k_B T}} dx } = \frac{1}{Z_a} \left\langle x^2 e^{-\frac{V(x)}{k_B T}} \right\rangle_a \, , \label{eq:msd-nqe-ensemble}
    \end{eqnarray}
    where we define the partition function of a single site $Z_a \equiv \left\langle e^{-{V(x)}/{k_B T}} \right\rangle_a$. We highlight that Eq.\,(\ref{eq:msd-nqe-ensemble}) is valid only for times shorter than the escape time while for sufficiently long times, the moments will behave as Eq.\,(\ref{eq:msd-long-time}), as we can see in Fig. \ref{fig:moments}(a).
    }
    
    \subsection{Periodic observables}
	
	For observables that are periodic with the spacing $a$, such as the internal energy observable $E \equiv \lim_{t \to \infty} \langle V(x) \rangle_t$, we have that
	\begin{eqnarray}
		E  &=& \lim_{t \to \infty} \int_{-\infty}^\infty V(x)  \frac{e^{-\frac{V(x)}{k_B T}} e^{-\frac{x^2}{4 D^* t}}}{\sqrt{4 \pi D t}} dx \, , \label{eq:internal-energy}
	\end{eqnarray}
	where we have used $P_t$ in Eq.\,(\ref{eq:Pt-theoretical-0}), as the asymptotic correction in Eq.\,(\ref{eq:Pt-theoretical}) will yield $O(t^{-1})$ contributions. 
	Once again, we make use of the scaling $y \equiv x/\sqrt{t}$ to write
	\begin{eqnarray}
		\langle V(x) \rangle_t &\approx& \int_{-\infty}^\infty \frac{V(y \sqrt{t}) e^{-\frac{V(y \sqrt{t})}{k_B T}}}{\left\langle e^{-\frac{V(x)}{k_B T}} \right\rangle_a}   \frac{e^{-\frac{y^2}{4 D^* }}}{\sqrt{4 \pi D^* }} dy \, .
	\end{eqnarray}	
	In the long-time limit, the term $V(y \sqrt{t}) e^{-{V(y \sqrt{t})}/{k_B T}}$ will oscillate rapidly, which allows us to replace its value for an average in a period, that is
	\begin{eqnarray}
			\langle V(x) \rangle_t &\approx& \frac{\left\langle V(x) e^{-\frac{V(x)}{k_B T}} \right\rangle_a}{\left\langle e^{-\frac{V(x)}{k_B T}} \right\rangle_a}  \int_{-\infty}^\infty  \frac{e^{-\frac{x^2}{4 \pi D^* t}} }{\sqrt{4 \pi D^* t}} dx \, , \label{eq:Vt_ensemble}
	\end{eqnarray}
	where the integral on the right-hand side is clearly unity and the denominator is the partition function of a single site $Z_a \equiv \left\langle e^{-{V(x)}/{k_B T}} \right\rangle_a$. We conclude that the internal energy observable converges to the expected result of thermal equilibrium in a single cell, that is,
	\begin{eqnarray}
	     E & = & \frac{1}{Z_a} {\left\langle V(x) e^{-\frac{V(x)}{k_B T}} \right\rangle_a}  \, . \label{eq:prediction-Ei}
	\end{eqnarray}
	\new{Unlike the similar Eq.\,(\ref{eq:msd-nqe-ensemble}), which is only valid for short times, Eq.\,(\ref{eq:prediction-Ei}) is valid in the long time limit and represents a true stationary-like state. Further, Eq.\,(\ref{eq:msd-nqe-ensemble}) holds for deep wells or low temperatures, $v_0/k_B T \gg 1$, while Eq.\,(\ref{eq:prediction-Ei}) has a general validity. We show the validity of Eq.\,(\ref{eq:prediction-Ei}) in Fig. \ref{fig:ensemble-observables}(a).}
	In the case of the cosine potential in Eq.\,(\ref{eq:cosine-pot-def}) is $E = - (V_0/2) I_1\left({V_0}/{2 k_B T}\right)/I_0\left({V_0}/{2 k_B T}\right)$, where $I_n(x)$ is the $n$-th modified Bessel function of the first kind. This can be extended for any $a$-periodic observable $\Ob (x)$,
	\begin{eqnarray}
	   \lim_{t \to \infty} \langle \Ob(x) \rangle_t = \frac{1}{Z_a} {\left\langle \Ob(x) e^{-\frac{V(x)}{k_B T}} \right\rangle_a} \, . \label{eq:lattice-eq}
	\end{eqnarray}
	This is clearly very similar to standard canonical averaging, found for usual confining systems. 
	
	\subsection{Integrable observables}
	As mentioned, for systems with a non-binding potential \cite{Aghion2019,Aghion2020}, a form of non-normalizable Boltzmann-Gibbs statistics emerges. There is a class of observables $\mathcal{O}(x)$ that is integrable with respect to the infinite density, that is, $\lim_{t \to \infty} \int_{-\infty}^{\infty} {\cal Z}_t P_t(x) {\cal O}(x) dx = \int_{-\infty}^{\infty} e^{- {V(x)}/{k_B T}} {\cal O}(x) dx < \infty $. 
	In the long-time limit, we use Eq.\,(\ref{eq:infinite-density-def}) to write the PDF as $P_t(x) \approx e^{- {V(x)}/{k_B T}}/\Z$, and the ensemble average can be calculated as
	\begin{eqnarray}
		\langle \mathcal{O} (x) \rangle_t & \sim & \frac{1}{\mathcal{Z}_t} \int_{-\infty}^{\infty} \mathcal{O} (x) e^{-\frac{V(x)}{k_B T}} dx  \label{eq:infinite-avg} \, .
	\end{eqnarray}
	An example is the indicator function, defined as
	\begin{eqnarray}
		\Theta (x) = \left\{\begin{array}{l}
			1 ~ \text{for} ~ x_A < x < x_B \\
			0 ~ \text{otherwise}
		\end{array} \right. \, , \label{eq:indicator-func}
	\end{eqnarray}
	with an ensemble average
	\begin{eqnarray}
		\langle \Theta (x) \rangle_t & \sim & \frac{1}{\mathcal{Z}_t} \int_{x_A}^{x_B} e^{-\frac{V(x)}{k_B T}} dx  \label{eq:occ1-theo} \, .
	\end{eqnarray}
    Observables that are integrable with respect to the \new{time-invariant infinite density} do not follow regular ergodicity, as we will see in Sec. \ref{sec:ergodicity}. In panel (b) of Fig.\,\ref{fig:ensemble-observables}, we plot a comparison between the numerical ensemble average and the long-time approximation,  Eq.\,(\ref{eq:occ1-theo}), of the indicator function for $x_A = a/10$ and $x_B = a/5$. 

    We see that periodic observables (such as the energy) and non-integrable observables (such as the indicator function) are sensitive to the fine structure of the density, while the positional moments are not. The averages of integrable observables depend on $D^*t$ through $\Z$, while the periodic observables do not. Generally, the observables are functionals of the path $x_\eta(t)$, that is, $O(x_\eta(t))$. For the \new{indicator} function, this observable is zero most of the time, with long power law distributed times between return events while the energy observable is non-zero nearly all the time, hence the two observables have vastly different behaviors.
	
	\subsection{The virial observable}
	
\begin{figure}
        \includegraphics[width = 0.45\textwidth]{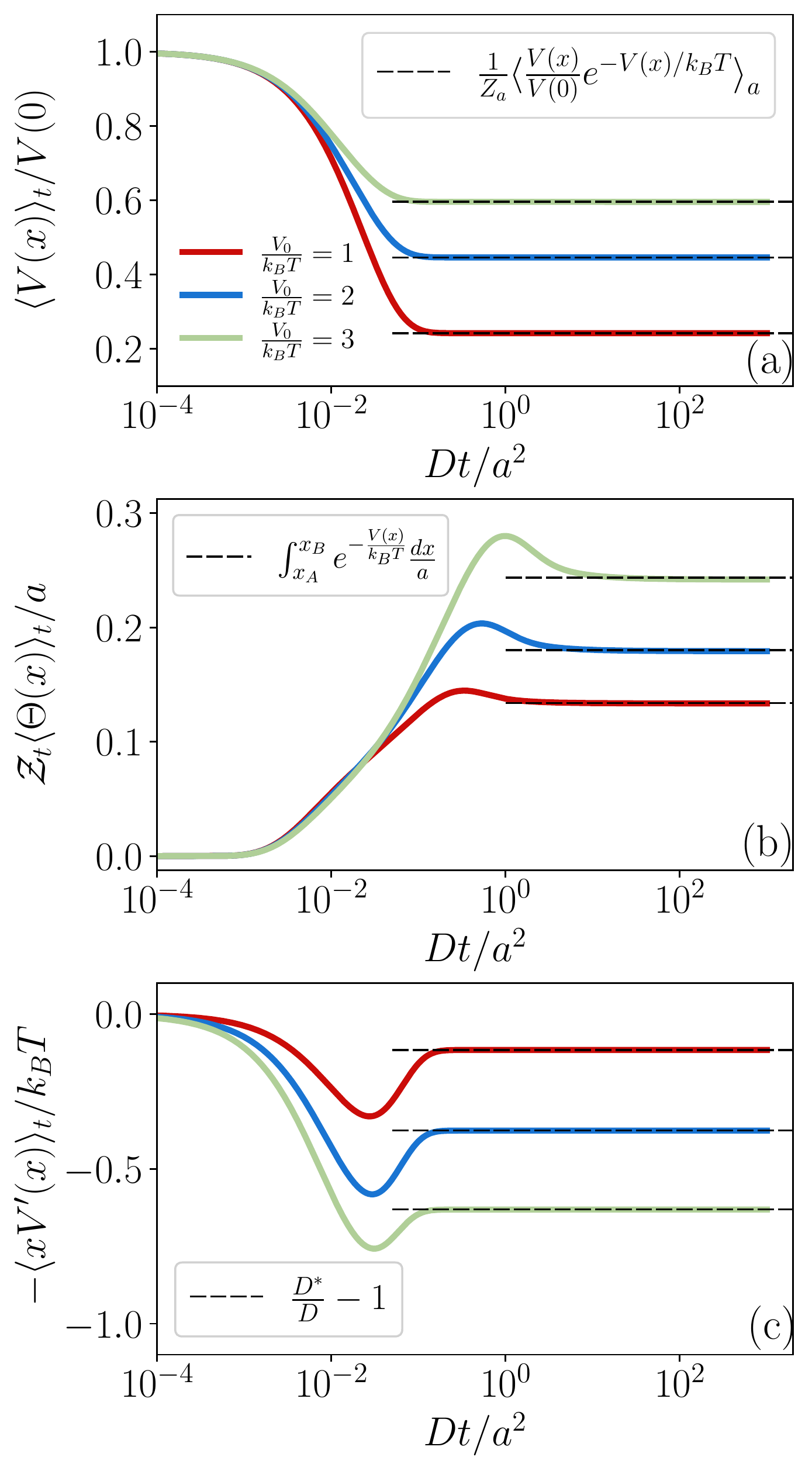}
		\caption{ \new{The ensemble average of three types of observables (solid lines) versus time: (a) the internal energy, (b) the indicator function with $x_A=a/10$ and $x_B=a/5$ and (c) the virial observable, for different values of $V_0/k_B T$ shown in the legend of panel (a). 
        The dashed black lines represent our long-time predictions (shown in the framed legends): (a) Eq. (\ref{eq:prediction-Ei}), (b) Eq. (\ref{eq:occ1-theo}) and (c) Eq. (\ref{eq:virial-theo}), and they indeed show excellent agreement in the long time limit. 
        The observables in panels (a) and (b) clearly depend on the Boltzmann-Gibbs factor $e^{-V(x)/k_B T}$, while the virial observable in panel (c) does not.} \label{fig:ensemble-observables}}
	\end{figure}

	We saw in Eq.\,(\ref{eq:lattice-eq}) and in Eq.\,(\ref{eq:infinite-avg}) how the Boltzmann-Gibbs factor is used to obtain statistical information on the system. Therefore, it is natural to wonder how thermodynamic relations hold for this system. Thus, we will now study the virial theorem. 
	The average of the observable related to the virial theorem $- x V'(x)$ can also be calculated through Eq.\,(\ref{eq:Pt-theoretical}). This observable consists of an oscillating function whose amplitude increases linearly with the position,
	\begin{eqnarray}
			\langle - x V'(x) \rangle_t &\approx& -  \int_{-\infty}^\infty  x V'(x)   \frac{e^{-\frac{V(x)}{k_B T}-\frac{x^2}{4 D^* t}}}{ \Z } \nonumber \\
			& ~ & ~~~~~~ \left[ 1 - \frac{x U_1(x)}{2D^*t} \right] dx \, , 
	\end{eqnarray} 
	where, since $\langle V'(x)  e^{-{V(x)}/{k_B T}}  \rangle_a = 0$, the leading term is null, so we must look to the first correction in time. Using the definition of $U_1(x)$ in Eq. (\ref{eq:U1-def}), we obtain that
	\begin{eqnarray}
	    \left\langle V'(x) e^{- \frac{V(x)}{k_B T}} U_1(x) \right\rangle_a &\approx& \frac{1}{a} \int_{0}^a \frac{\partial V(x)}{\partial x} e^{-\frac{V(x)}{k_B T}} U_1(x) dx  \nonumber \\
        &\approx& \frac{k_B T}{a} \int_{0}^a e^{-\frac{V(x)}{k_B T}} U_1'(x) dx  \nonumber \\
        &\approx& k_B T \left[\sqrt{\frac{D^*}{D}} - \sqrt{\frac{D}{D^*}} \right] \, ,
	\end{eqnarray}
	and the virial observable becomes,
	\begin{eqnarray}
	    \langle - x V'(x) \rangle_t &\approx& k_B T \left[\sqrt{\frac{D^*}{D}} - \sqrt{\frac{D}{D^*}} \right] \int_{-\infty}^{\infty} \frac{x^2 e^{-\frac{x^2}{4 D^* t}}}{2 D^* t} \frac{dx}{\Z} \nonumber \\
	   &\approx& - k_B T  \left[ 1 - \frac{D^*}{D} \right] \label{eq:virial-theo} \, .
	\end{eqnarray}
	For systems with confining potentials, there is no diffusion at long times, that is $ d \langle x^2 \rangle / dt= 2 D^* = 0$, and therefore $\langle - x V'(x) \rangle_t = - k_B T$, and we recover, as expected, the regular virial theorem. In the opposite limit of free diffusion, $D^* = D$, and then, as expected, the right-hand side of Eq.\,(\ref{eq:virial-theo}) gives zero.
	We compare our long-time prediction with the numerical calculation of the ensemble averages in Fig.\,\ref{fig:ensemble-observables}(c). Eq.\,(\ref{eq:virial-theo}) indicates that thermodynamic relations, like the virial theorem, can be extended to the study of Brownian motion in non-confining periodic systems, and below we continue with this theme, namely, extending the domain of the standard machinery of statistical mechanics.
	
	\section{Time averages and ergodicity \label{sec:ergodicity}}
	
	We may also study the system on the level of individual realizations, as described by the Langevin Equation\,(\ref{eq:langevin}). This corresponds to single particle trajectories as found for example in single-molecule experiments. 
    For each realization, there is a stochastic trajectory $x_\eta(t)$, and observables which are functions of the position, that is, $\mathcal{O}_\eta \equiv \mathcal{O}(x_\eta)$, are also stochastic variables. 
    As usual, as the number of realizations $\mathcal{N}$ becomes large, averaging over these trajectories will converge to the expectation obtained using $P_t(x)$.
	
	As in experimental settings it may be impractical to reproduce the experiment sufficient times to obtain the ensemble averages, we may look instead at the time averages of observables, defined as
	\begin{eqnarray}
		\overline{\mathcal{O}}_t = \frac{1}{t} \int_{0}^{t} {\cal O}\big(x_\eta(t') \big) dt' \label{eq:time-ob} \, ,
	\end{eqnarray}
	for each trajectory $x_\eta$. These time averages are also stochastic variables and for ergodic systems, we will have that the time averages will converge, at very long times, to the ensemble averages, that is, $ \lim_{t \to \infty} \overline{\mathcal{O}}_t/ \langle \mathcal{O}(x) \rangle_t = 1$.
	
	\subsection{Positional moments}
	
	In the case of the ensemble averages of the mean square displacement $\langle x^2 \rangle_t$, even though this observable is clearly describing a non-equilibrium feature of the system, the increments of the position are stationary. We can define the displacement \new{over a time interval $\Delta t$ as $\delta x(t) \equiv x(t+\Delta t) - x(t)$}, with the time-averaged MSD being \cite{Metzler2014}
	\begin{eqnarray}
	    \overline{\delta x^2}_t &=& \frac{1}{t - \Delta t} \int_0^{t-\Delta t} \delta x^2(t')\, dt' \, . \label{eq:msd-time}
	\end{eqnarray}
	If the height of the potential barrier is much larger than the temperature, for times $\Delta t$ much shorter than the escape time, the particle will be in thermal equilibrium within a single well, and we have
	\begin{eqnarray}
	    \overline{\delta x^2}_t & \approx & \frac{1}{t - \Delta t} \int_0^{t-\Delta t} \left( x^2(t'+\Delta t) + x^2(t') \right) dt' = \nonumber \\
	    &\approx & \frac{2}{Z_a} \left\langle x^2 e^{-\frac{V(x)}{k_B T}} \right\rangle_a \, , \label{eq:msd-nqe}
	\end{eqnarray}
	which resembles Eq.\,(\ref{eq:msd-nqe-ensemble}), with a factor of 2, as previously seen in \cite{Cherstvy2018}.
	We compare the numerical evaluation of $\overline{\delta x^2}_t$, using a Langevin equation, for a single trajectory with our predictions in panel (b) of Fig.\,(\ref{fig:moments}). We see that for short $\Delta t$, Eq.\,(\ref{eq:msd-nqe}) holds while for longer time scales it converges to Eq.\,(\ref{eq:msd}), namely,
    \begin{eqnarray}
        \overline{\delta x^2}_t \rightarrow 2 D^* \Delta t \, . \label{eq:msd-long-time}
    \end{eqnarray}
    This means that in the long-time limit, namely, when the measurement time $t$ is larger than $\Delta t$ and also much larger than the escape time from a well, we get standard, though non-equilibrium, ergodic behavior for the mean square displacement (by non-equilibrium we mean that the observable is determined by $D^*$ and is not related to the Boltzmann-Gibbs factor).

	\subsection{Periodic observables \label{subsec:periodic-observables-time-average}}
	
	We showed in the previous Section that the ensemble averages of periodic observables are equivalent to a Boltzmann-Gibbs average over a period of the potential. If the observable is ergodic, then we must have that, for very long times, the time averages of a single realization will converge to the ensemble averages. To show that this is indeed the case, we will study the statistics of the time averages of the energy observable,
	\begin{eqnarray}
	    \overline{V}_t &=& \frac{1}{t} \int_{0}^{t} V \big(x_\eta(t') \big) dt' \, , \label{eq:energy-time}
	\end{eqnarray}
	obtained using Eq.\,(\ref{eq:time-ob}).	As we mentioned, the ergodicity of this observable means
	\begin{eqnarray}
	    \lim_{t \to \infty} \overline{V}_t = E \label{eq:ergod-energy} \, ,
	\end{eqnarray}
	where the right-hand side is given by Eq.\,(\ref{eq:prediction-Ei}). To show that the energy observable follows ergodicity, that is, Eq.\,(\ref{eq:ergod-energy}), we must first ensure that, for long times, we have
	\begin{eqnarray}
	   \lim_{t \to \infty} \langle \overline{V}_t \rangle &=& \lim_{t \to \infty} \frac{1}{t} \int_0^t \langle V(x(t')) \rangle_{t'} dt' = E \, , \label{eq:connection-averages}
	\end{eqnarray}
	which is clearly true as $\langle V(x) \rangle_t$ is constant in time, as seen in Eq.\,(\ref{eq:Vt_ensemble}). From $\overline{V}_t$, we define $\Delta \overline{V}_t \equiv \overline{V}_t - \langle \overline{V}_t \rangle$,
	\begin{eqnarray}
	    \Delta \overline{V}_t &=&  \frac{1}{t} \int_{0}^{t} V \big(x_\eta(t') \big) dt' - \langle \overline{V}_t \rangle \nonumber \\
	    &=&  \frac{1}{t} \int_{0}^{t} \left[ V \big(x_\eta(t') \big) - \langle \overline{V}_t \rangle \right] dt'  \, ,
	\end{eqnarray}	
    where we remark that the term $\langle \overline{V}_t \rangle$, which in the last line we placed inside the integral, only depends on the final time $t$ and is therefore constant through the integration. This allows us to define, for each time $t$, the function $\Delta V(t') \equiv V(x(t')) - \langle \overline{V}_t \rangle$.
    
    The variance of $\overline{V}_t$ is given by
    \begin{eqnarray}
        \left\langle \Delta \overline{V}_t^2 \right\rangle &=& \left\langle \left( \frac{1}{t} \int_0^t \Delta V(t_1) dt_1 \right) \left( \frac{1}{t} \int_0^t \Delta V(t_2) dt_2 \right) \right\rangle \nonumber \\
        &=& \frac{1}{t^2} \int_{0}^{t}\int_{0}^{t} dt_1 dt_2 \left\langle \Delta V(t_1) \Delta V(t_2) \right\rangle \, , \label{eq:variance-mean}
	\end{eqnarray}
	where $\langle \Delta V(t_1) \Delta V(t_2) \rangle$ is the correlation function. We will show that for long times $\langle  \Delta\overline{ V }^2 \rangle_t \to 0$, which, combined with Eq.\,(\ref{eq:connection-averages}), ensures that the energy observable exhibits ergodic features. 
	
	For times much larger than the escape time, in Eq.\,(\ref{eq:Pt-theoretical}) we see that for values of $x \ll \sqrt{2 D^* t}$ (the diffusive length scale), the Gaussian contribution of the PDF is approximately constant, and the PDF itself is proportional to the Boltzmann-Gibbs factor. Let us replace the cutoff of the Gaussian with a sharp cutoff by placing the system in a $2L$-sized box, with reflecting boundaries, where $L \gg a$. For simplicity, let us also consider that $L = n a$, with $n$ integer. For this confined system, in the long-time limit, Eq.\,(\ref{eq:variance-mean}) can be written as \cite{Dechant2011}
    \begin{eqnarray}
			\left\langle \Delta \overline{V}_t^2 \right\rangle & \approx &  \frac{2}{D t} \int_{-na}^{na} dx \frac{e^{\frac{V (x)}{k_B T}}}{Z_L} \left[ \int_{x}^{na} \Delta V(y)  e^{- \frac{V(y)}{k_B T}} dy \right]^2 \, ,  \nonumber \\ \label{eq:var-with-L}
	\end{eqnarray}
	where the partition function in the denominator can be written as $Z_L = \int_{-na}^{na} e^{- {V(x)}/{k_B T}} dx = 2 n a Z_a$. Using that $\int_{x}^{x+a} \Delta V(y)  e^{- {V(y)}/{k_B T}} dy = 0$, and the periodicity of the potential, we obtain that
	\begin{eqnarray}
	    \langle \Delta \overline{V}_t^2 \rangle & \approx &  \frac{2}{D t} (2n) \int_{0}^{a} dx \frac{e^{\frac{V (x)}{k_B T}}}{2 n a Z_a} \left[ \int_{x}^{a} \Delta V(y)  e^{- \frac{V(y)}{k_B T}} dy \right]^2 \nonumber \\
	    & \approx & \frac{2}{Dt} \int_{0}^{a} \frac{dx}{a} \frac{e^{\frac{V (x)}{k_B T}}}{Z_a} \left[ \int_{x}^{a} \Delta V(y)  e^{- \frac{V(y)}{k_B T}} dy \right]^2  \, ,  \label{eq:var-energy}
	\end{eqnarray}
    a result valid for all periodic observables. Notice that Eq.\,(\ref{eq:var-energy}) does not depend on the auxiliary lengthscale $L$, which was used in Eq.\,(\ref{eq:var-with-L}) as a tool only. The PDF$(\overline{V}_t)$ converges to a Gaussian with variance given by Eq.\,(\ref{eq:var-energy}), which decreases in time.    This is a feature present in equilibrium systems where, for a very long time ($t \to \infty$), the time average of a single realization will converge to the ensemble average. We have verified numerically the validity of Eq.\,(\ref{eq:var-energy}), integrating the Langevin Eq.\,(\ref{eq:langevin}). We compare the numerical results with our predictions in panel (a) of figure\,\ref{fig:observables-ergodic}.
    
    \subsection{Integrable observables}
    
	In the long-time limit, the ensemble average of the time average of an observable that is integrable with respect to the infinite density, namely, an observable that satisfies $\int_{-\infty}^\infty \mathcal{O}(x) e^{- {V(x)}/{k_B T}} dx < \infty$, is calculated to be
	\begin{eqnarray}
	\langle \overline{{\cal O}}_t \rangle & \approx & \frac{1}{t} \int_{0}^{t} \langle {{\cal O}}(x) \rangle_{{t'}} dt' \nonumber \\
        &\approx&  \int_{0}^{t} \frac{dt'}{t \mathcal{Z}_{t'}} \int_{-\infty}^{\infty} {\cal O}(x) e^{-\frac{V(x)}{k_B T}} dx \nonumber \\ 
	& \approx & 2 \langle {\cal O}(x) \rangle_t \label{eq:relation-between-averages}  \, .
\end{eqnarray}
    The doubling effect we see in Eq.\,(\ref{eq:relation-between-averages}) is related to the time integral over $t^{-1/2}$, it appears also for other related problems (see \cite{Aghion2019,Aghion2020}). Eq.\,(\ref{eq:relation-between-averages}) is a relation between the ensemble average of the time average and the ensemble average of the observable in Eq.\,(\ref{eq:occ1-theo}). We now briefly discuss the time average, focusing on a particular observable, the indicator function $\Theta(x)$, defined in Eq.\,(\ref{eq:indicator-func}).

	At the level of individual trajectories, the time average of the indicator function $\Theta(x)$ is equivalent to the occupation time the particle spends inside the interval $(x_A,x_B)$ divided by the measurement time, which is a random variable in the range $(0,1)$.  For usual ergodic systems, such as a Brownian Particle in a confining harmonic potential, this time average in the long-time limit will approach the probability of being in that interval, that is,
	\begin{eqnarray}
	    \lim_{t\to \infty} \overline{\Theta}_t = \frac{1}{Z} \int_{x_A}^{x_B} e^{-V(x)/k_B T} dx   \, ,
	\end{eqnarray}
	where $Z = \int_{-\infty}^\infty e^{- \frac{V(y)}{k_B T}} dy$ is the usual partition function.
	In our case, fluctuations of the time averages of integrable observables remain non-trivial, unlike the energy and the mean square displacement considered so far.

    \begin{figure}
        \includegraphics[width = 0.45\textwidth]{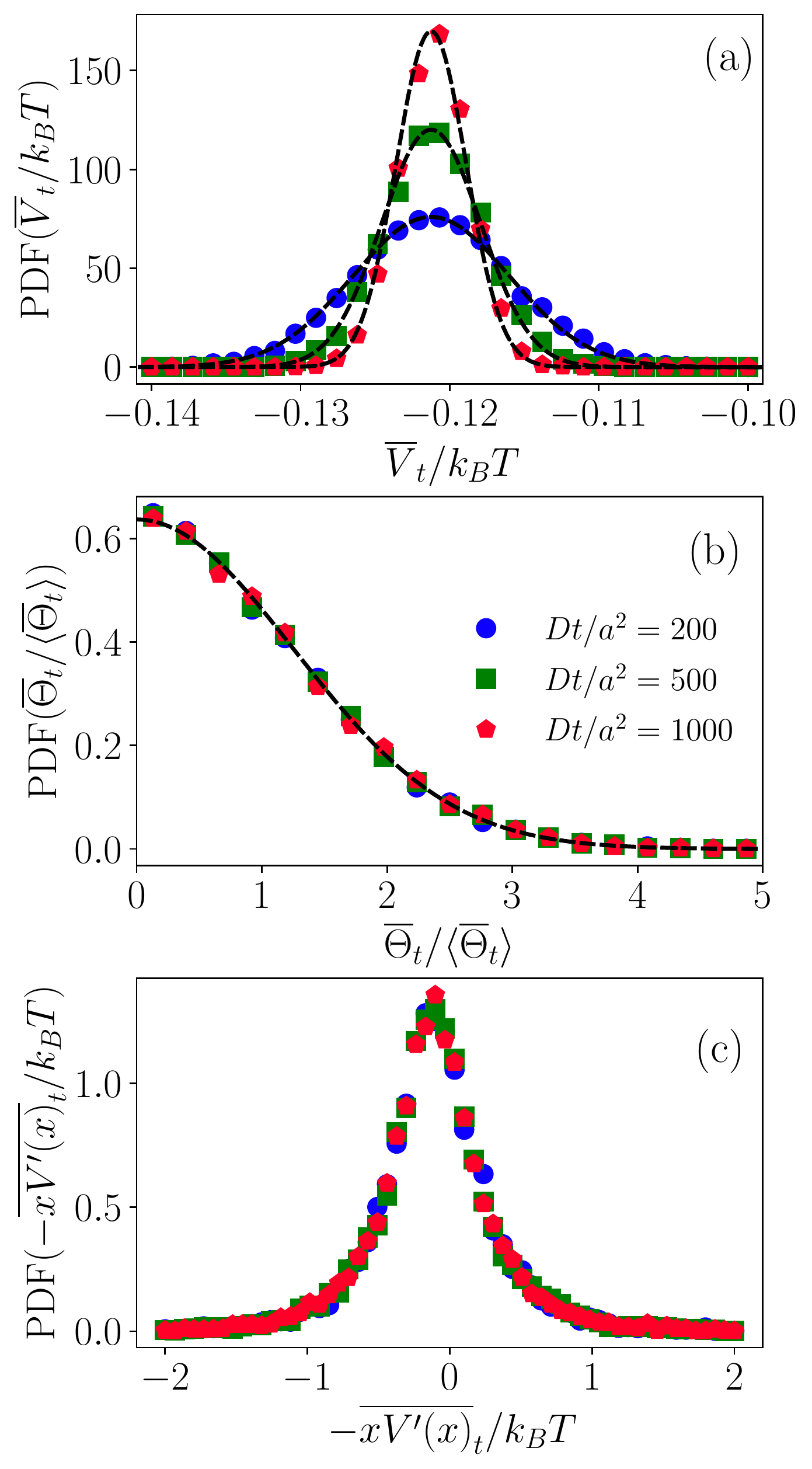}
		\caption{\new{The PDF of the time average of three different observables (symbols): (a) the internal energy, (b) the ratio between time- and ensemble average of the indicator function with $x_A=-a/2$ and $x_B = a/2$, and (c) the virial observable, for three different measurement times shown in the legend of panel (b).
        For long times, the PDF of the internal energy approaches a normal distribution (dashed lines in panel (a)), with the mean given by Eq. (\ref{eq:prediction-Ei}), and the variance given by Eq.\,(\ref{eq:var-energy}). We see that as we increase measurement time, the distribution approaches a narrow delta function, namely, the energy observable is perfectly ergodic, even though the standard normalization is not found in our system.
        The statistics of the ratio between time and ensemble averages of the indicator function approaches that of half a Gaussian (dashed lines in panel (b)) as predicted in Eq.\,(\ref{eq:return-time-statistics}), which is a manifestation of the Darling-Kac theorem for integrable observables, as mentioned in the text. In this case, unlike the internal energy, the distribution is time-invariant when the measurement time is long.
        We can also see that the virial observable is not ergodic, since the distribution becomes time-invariant, as seen in panel (c), similar to the PDF in panel (b).} \label{fig:observables-ergodic}}
	\end{figure}
	\begin{figure}
	    \centering
	    \includegraphics[width=0.475\textwidth]{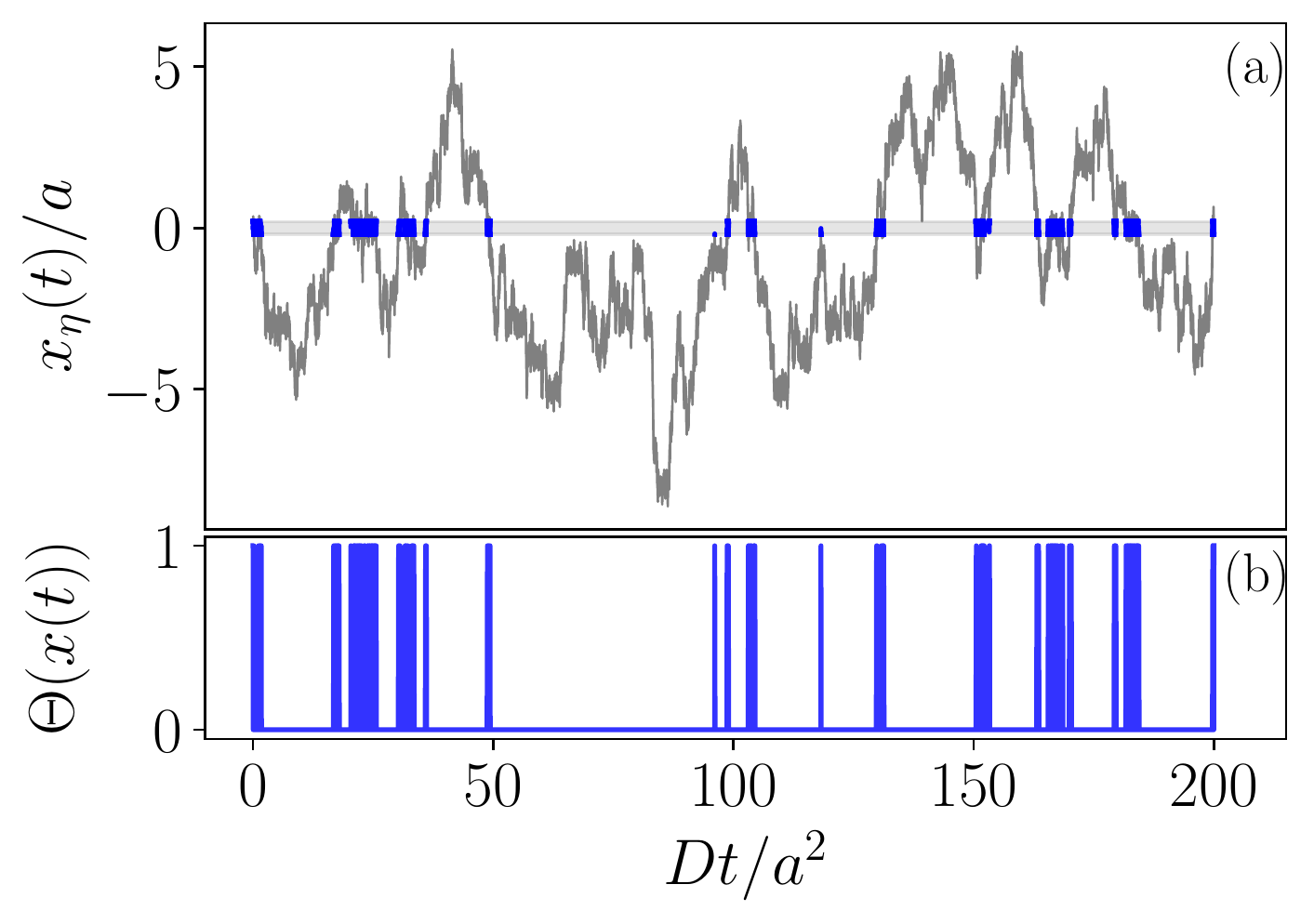}
	    \caption{\new{Panel (a): the trajectory of a single realization for $V_0/k_B T = 2$. Panel (b): the indicator function for the trajectory in panel (a), with $x_A = -a/5$ and $x_B = a/5$, the blue highlight indicates when the particle is inside the interval. 
        The fluctuations of the fat tailed distributed times spent inside the interval are small compared to the time the particle spends outside  and therefore, the statistics of the return time control the fluctuations.}}
	    \label{fig:trajectory}
	\end{figure}
	
	We observed numerically that the distribution for the first return time $\tau$ is  fat-tailed, where for large $\tau$ we have the power-law $\tau^{-3/2}$ \cite{Aghion2019,Aghion2020}, a result expected for Brownian motion \cite{Godreche2001}. We define the ratio
	\begin{eqnarray}
	    \xi = \frac{\overline{\Theta}_t}{\langle \overline{\Theta}_t \rangle} \, ,
	\end{eqnarray}
	which is a random variable with unit mean. It gives the ratio of the total time the particle spends in the domain $(x_A,x_B)$ in this realization and the mean of the same observable. In Fig.\,(\ref{fig:trajectory}) we plot the trajectory of a particle, highlighting the times the particle spends inside the domain. The statistics of the return times will control the fluctuations of this ratio, and it approaches the ratio between the number $n$ of crossings into the region in this realization and the average number of such crossings $\langle n \rangle$, that is $\xi = n/\langle n \rangle$. Using renewal theory, we obtain the PDF \cite{Godreche2001}
    \begin{eqnarray}
        \mathrm{PDF}(\xi) = \frac{2}{\pi} e^{-\frac{\xi^2}{\pi}} \, , \label{eq:return-time-statistics}
    \end{eqnarray}
    which is equivalent to half a Gaussian, as $\xi \geq 0$.
    This result is a manifestation of the Darling-Kac theorem. In panel (b) of Fig.\,(\ref{fig:observables-ergodic}) we compare Eq.\,(\ref{eq:return-time-statistics}) with numerical results obtained using the Langevin Equation\,(\ref{eq:langevin}). For other physical applications of the Darling-Kac theorem see \cite{Korabel2012,Barkai2021}.
	
	\subsection{The virial observable}
	
	Since the ensemble average of the virial observable, Eq.\,(\ref{eq:virial-theo}), does not depend on time, unlike what we have found for integrable observables such as the indicator function $\Theta(x)$, the expectation of time averages will converge to the ensemble averages, that is
	\begin{eqnarray}
	    \langle -\overline{x V'(x)}_t \rangle = \langle - x V'(x) \rangle_t \, .
	\end{eqnarray}
	This does not mean that the virial observable is ergodic. As we see in panel (c) of Fig. \ref{fig:observables-ergodic}, the variance of the PDF of the time average of the virial observable does not decrease when we increase the duration of the time average, and therefore a single realization is never sufficient to accurately obtain the \new{ensemble} average.
	
\begin{figure}
	    \centering
	    \includegraphics[width=0.475\textwidth]{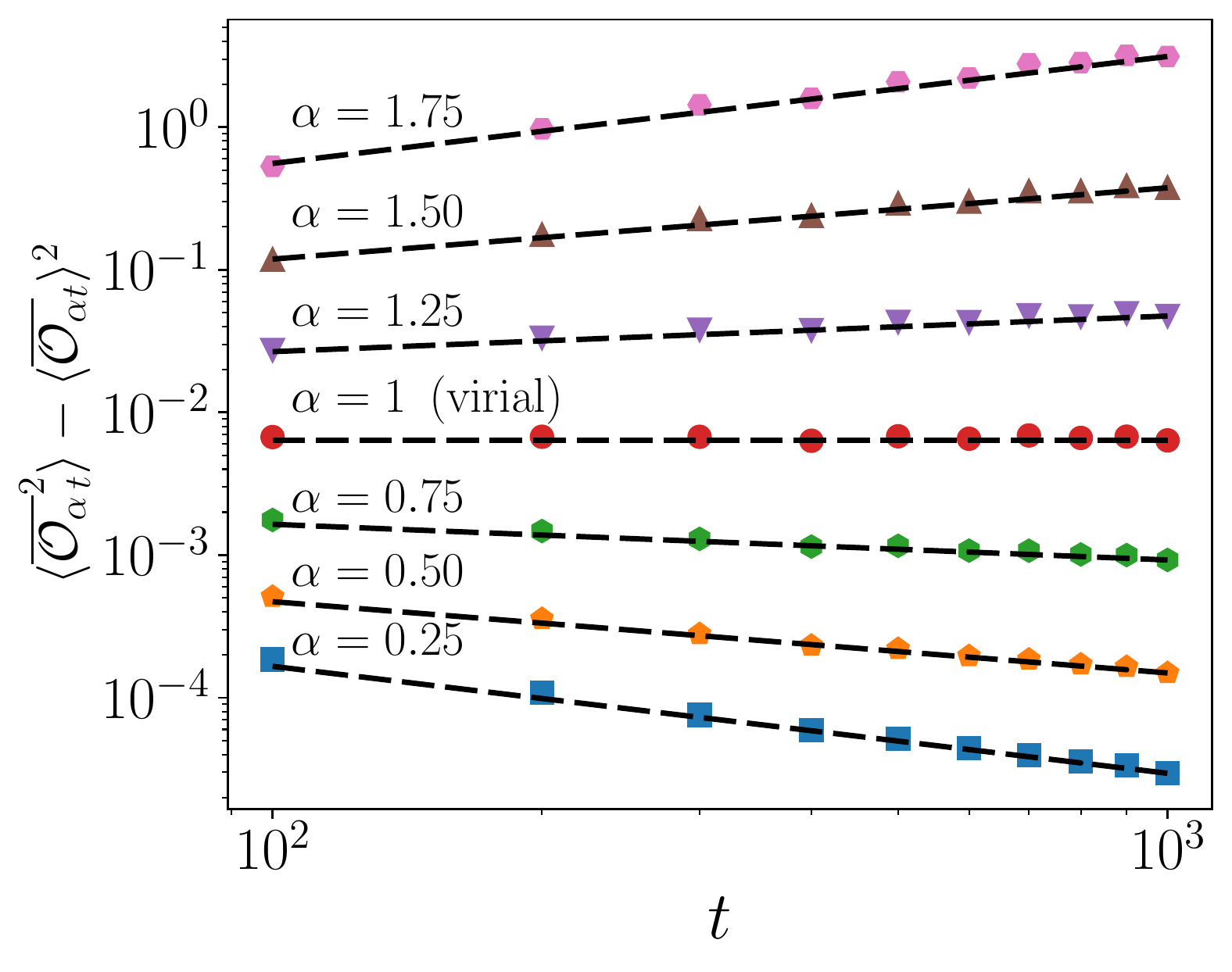}
	    \caption{\new{The squared variance of the time-average of the observable $\mathcal{O}_\alpha$ (symbols), Eq.\,(\ref{eq:O-alpha}), with a fit proportional to $t^{\alpha-1}$ (dashed lines). Note that the only time-independent variance is for $\alpha = 1$, which is equivalent to the virial. We have used $V_0/k_B T = 1$.}\label{fig:variances}}
	    
	\end{figure}

	\new{
	The virial is a special observable as it consists of the product of an observable that is insensitive to the fine structure, $x$, and a periodic observable, the force $-V’(x)$. Consider the following (purely mathematical) observable $\mathcal{O}_2(x) = (x/a)^2 \cos (2 \pi x/a)$. The ensemble average can be calculated using $P_t(x)$ in  Eq.\,(\ref{eq:Pt-theoretical-0}) as
	\begin{eqnarray}
	   \langle \mathcal{O}_2 \rangle_t &\approx&  \int_{-\infty}^\infty \left( \frac{x}{a} \right)^2 \cos \left( \frac{2 \pi x}{a} \right)  \frac{e^{-\frac{V(x)}{k_B T}} e^{-\frac{x^2}{4 D^* t}}}{\sqrt{4 \pi D t}} dx \nonumber \\
	   & \approx & \frac{\left\langle \cos \left( \frac{2 \pi x}{a} \right) e^{-\frac{V(x)}{k_B T}} \right\rangle_a}{\left\langle e^{-\frac{V(x)}{k_B T}} \right\rangle_a}  \int_{-\infty}^\infty \frac{x^2}{a^2} \frac{e^{-\frac{x^2}{4 \pi D^* t}} }{\sqrt{4 \pi D^* t}} dx \nonumber \\
	   & \approx & \frac{\left\langle \cos \left( \frac{2 \pi x}{a} \right) e^{-\frac{V(x)}{k_B T}} \right\rangle_a}{\left\langle e^{-\frac{V(x)}{k_B T}} \right\rangle_a} \left( \frac{2 D^* t}{a^2} \right) \,  , \label{eq:O2-ensemble-avg}
	\end{eqnarray}
	where we have used the same scaling arguments as in Eq.\,(\ref{eq:Vt_ensemble}) to replace the oscillating terms by their average in a unit cell. Note that, unlike what we see for the virial observable, the time-dependence of the mean is controlled entirely by $x^2$, with the only contribution of the periodic term being a multiplicative constant. This indicates that the first ingredient to obtaining stationary time averages is that the oscillating function must have zero mean in a unit cell. We already saw that the virial observable follows this restriction, as $\langle V'(x)  e^{-{V(x)}/{k_B T}}  \rangle_a = 0$, and the further corrections of $P_t(x)$, present in Eq.\,(\ref{eq:Pt-theoretical}), become necessary.
	
	We now define a different family of even observable as
	\begin{eqnarray}
	    \mathcal{O}_\alpha (x) = 
			\mathrm{sgn}(x) (|x|/a)^\alpha \sin(2 \pi x/a) 
		 \, , \label{eq:O-alpha}
	\end{eqnarray}
	where $\mathrm{sgn}(x)$ is the sign function. We will focus on the potential of Eq.\,(\ref{eq:cosine-pot-def}), so that the force $-a V'(x)/V_0 = \sin (2 \pi x/a)$. The case $\alpha=1$, the observable is effectively the virial observable, for $\alpha = 0$, the observable is proportional to the force function while for any other value of $\alpha$, the observable is purely mathematical. Separately, the time-average of the periodic observable is ergodic (as we showed in Sec. \ref{subsec:periodic-observables-time-average}) with a variance that decreases as $t^{-1/2}$ and the variance of the time-average of the coarse-grained observable grows as $t^{\alpha/2}$. A very rough assumption, the validity of which we numerically show in Fig.\,\ref{fig:variances}, is to posit that the variance of the product will be proportional to the product of the individual variances, and therefore proportional to $t^{\alpha/2 - 1/2}$.
	
	For values of $\alpha < 1$, we observed in Fig.\,\ref{fig:variances} that the variance of $\overline{\mathcal{O}_\alpha}_t$ becomes narrower as time increases, similar to the internal energy (see Fig.\,\ref{fig:observables-ergodic}(a)) or more generally ergodic observables, while for $\alpha >1$, the PDF becomes broader. The virial observable, $\alpha = 1$, is a unique case where the variance of the time average (Fig. \ref{fig:variances}) and also the distribution (Fig. \ref{fig:observables-ergodic}(c)) become time-independent.} This surprising situation merits further study \new{as it shows that the virial observable is unique}.
	
	\subsection{Summary of Sec. \ref{sec:ergodicity}}
	
    We highlight that the observables studied in this Section can display vastly different ergodic properties. The ensemble averages of observables with the same periodicity as the potential, such as the internal energy, converge to those of a system in equilibrium in a single unit cell and display standard ergodicity properties. Other observables, such as the indicator function, have ensemble averages that are sensitive to the Boltzmann-Gibbs factor but do not follow regular equilibrium, as we see in Eq.\,(\ref{eq:occ1-theo}). The time averages of these observables do not follow regular ergodicity either, with their statistics being determined using the Darling-Kac theorem. \new{We also showed that the virial observable is non-trivial, as the distribution of the time average converges to a stationary PDF, see Figs. \ref{fig:observables-ergodic}(c) and \ref{fig:variances}.}
    
	\section{Entropy \label{sec:thermodynamics}}

	One important question is the behavior of the entropy of our system, given that the PDF in Eq.\,(\ref{eq:Pt-theoretical}) encompasses both the system's microscopic, i.e. on the length scale $a$, and macroscopic, found at the length scale $\sqrt{2 D^* t}$, behaviors. If one can acquire information about our system using bins of size $\Delta x$ that are much smaller than the lattice spacing, that is, $\Delta x \ll a$, the probability of finding a particle inside one of these bins is $p_t(x) \approx \Delta x P_t(x)$, and the entropy becomes
	\begin{eqnarray}
		\frac{S}{k_B} & = & - \sum_{x=-\infty}^{\infty} p_t(x)  \ln \left( p_t(x)  \right) \nonumber \\
		& \approx & - \int_{-\infty}^{\infty} P_t(x) \ln \left( P_t(x) \Delta x \right) dx \nonumber \\
		& \approx & \ln \frac{\Z}{\Delta x} + \frac{\langle V(x) \rangle_t}{k_B T} + \frac{\langle x^2 \rangle_t}{4 D^* t}\, , \label{eq:S-general}
	\end{eqnarray}
    where we have used that $\Delta x \ll a$ to replace the summation with an integral, and the last expression is valid at long times, with corrections of ${\cal O}(1/t)$. For the free particle case, $V(x) = 0$, the entropy can be calculated using Eq.\,(\ref{eq:S-general}) as
    \begin{eqnarray}
        \frac{S_0(D t)}{k_B} & \approx & \ln \frac{\sqrt{4 \pi D t}}{\Delta x} + \frac{1}{2} \label{eq:S-free-particle} \, ,
    \end{eqnarray}
    where the subscript 0 means that the periodic force is zero.
    Returning to the general case, we can use Eq.\,(\ref{eq:internal-energy}) to write $\langle V(x) \rangle_t/k_B T \approx E/k_B T$ and Eq.\,(\ref{eq:msd}) to write $\langle x^2 \rangle_t/4D^* t \approx 1/2$, obtaining an expression for the entropy of a system with a periodic potential $V(x)$ for large $t$ as
	\begin{eqnarray}
		\frac{S}{k_B} & \approx & \ln \frac{\sqrt{4 \pi D t}}{\Delta x} + \frac{1}{2} + \frac{E}{k_B T} \label{eq:S-theo-1} \, .
	\end{eqnarray}
    The corrections of order $O(t^{-1})$ to Eq.\,(\ref{eq:S-theo-1}) are dependent on the initial conditions of the system, while the leading terms, as we would expect from equilibrium, are independent of the initial conditions.
    Using that $D = D^* \left\langle e^{-{V}/{k_B T}} \right\rangle_a^2 = D^* Z_a^2$, we can write Eq.\,(\ref{eq:S-theo-1}) in an alternate, but equivalent, expression, 
	\begin{eqnarray}
		\frac{S}{k_B} & \approx & \ln \frac{\sqrt{4 \pi D^* t}}{\Delta x} + \frac{1}{2} + \frac{E - F_a}{k_B T} \, , \label{eq:S-theo-2}
	\end{eqnarray} 
	where we define the free energy in a single lattice period as $F_a = - k_B T \ln Z_a$. The first two terms in Eq.\,(\ref{eq:S-theo-1}) are equivalent to the entropy of a free particle, that is, $S_0(D t)$, while in Eq.\,(\ref{eq:S-theo-2}) the first two terms are equivalent to a free particle with a renormalized diffusion constant, $S_0(D^* t)$.
	
	It is also instructive to examine the opposite limit where the bin size is large, $\Delta x \gg a$, although still much smaller than the diffusive lengthscale $\Delta x \ll \sqrt{2 D^* t}$. The probability of finding the particle inside one of the bins is
	\begin{eqnarray}
	    p_t(j) \equiv \int_{(j-1/2)\Delta x}^{(j+1/2)\Delta x} P_t(x) dx \, , \label{eq:p-t}
	\end{eqnarray}
	where we integrate the PDF $P_t(x)$ around $x = j \Delta x$. The entropy, given by
	\begin{eqnarray}
	    \frac{S}{k_B} = - \sum_{j=-\infty}^{\infty} p_t(j) \ln p_t(j) \, ,
	\end{eqnarray}
	becomes a non-trivial function of the bin size.
	
    From our numerical observations (see Fig. \ref{fig:model-infiniteP}) and our theoretical predictions in Eq.\,(\ref{eq:Pt-theoretical}), the PDF contains a contribution from the fine structure of the lattice (the leading order contribution being the Boltzmann-Gibbs factor) and a coarse-grained contribution. We replace the PDF given by Eq.\,(\ref{eq:Pt-theoretical}) in Eq.\,(\ref{eq:p-t}) and change variables to the scaled variable $y \equiv x/\sqrt{t}$ (the effective bin size becoming $\delta y \equiv \Delta x/\sqrt{t}$) to write
    \begin{eqnarray}
        p_t(j) & \approx & \int_{(j-1/2)\delta y}^{(j+1/2)\delta y} e^{- \frac{V(\sqrt{t}y)}{k_B T}} e^{- \frac{y^2}{4D^*}} \frac{dy}{\sqrt{4\pi D}} \nonumber \\
        & \approx & \Delta x \frac{e^{-\frac{(j \Delta x)^2}{4 D^*t}}}{\sqrt{4 \pi D t}} \left\langle e^{- \frac{V}{k_B T}} \right\rangle_a  \nonumber \\ & \approx & \Delta x \frac{e^{-\frac{(j \Delta x)^2}{4 D^*t}}}{\sqrt{4 \pi D^* t}} \, .
    \end{eqnarray}
    where we have used that, for long times, as the limit of integration shrinks (since $\delta y = \Delta x/\sqrt{t}\to 0$), the Gaussian term is approximately constant in the region and as we are in the limit of bin size much larger than lattice spacing $a$, the Boltzmann-Gibbs factor oscillates several times in the domain, allowing us to replace it by its average. The last equation is obtained by replacing $D = D^* \left\langle e^{-{V}/{k_B T}} \right\rangle_a^2$. Finally, since $\Delta x \ll \sqrt{2 D^* t}$, we replace the summation with an integral and the entropy becomes
    \begin{eqnarray}
        \frac{S_{\mathrm{cg}}}{k_B} =  \ln \frac{\sqrt{4 \pi D^* t}}{\Delta x} + \frac{1}{2} = \frac{S_0(D^* t)}{k_B} \, , \label{eq:S-coarsed}
    \end{eqnarray}
    which is clearly different from Eqs.\,(\ref{eq:S-theo-1}) and (\ref{eq:S-theo-2}), as we no longer have the energy contributions $E$ and $F_a$ from the unit cell, leaving only the free particle entropy $S_0(D^* t)/k_B$.
    
    Since the entropy can only be defined up to a constant, we are typically interested in the difference between two entropies. Additionally, taking the difference of entropies can remove the time dependence in  Eqs.\,(\ref{eq:S-theo-1}), (\ref{eq:S-theo-2}) and\,(\ref{eq:S-coarsed}). We will focus in the following subsections on calculating different possible relative entropies.
	
	\subsection{Relative to a free-particle}
	
\begin{figure}
		\includegraphics[width = 0.475 \textwidth]{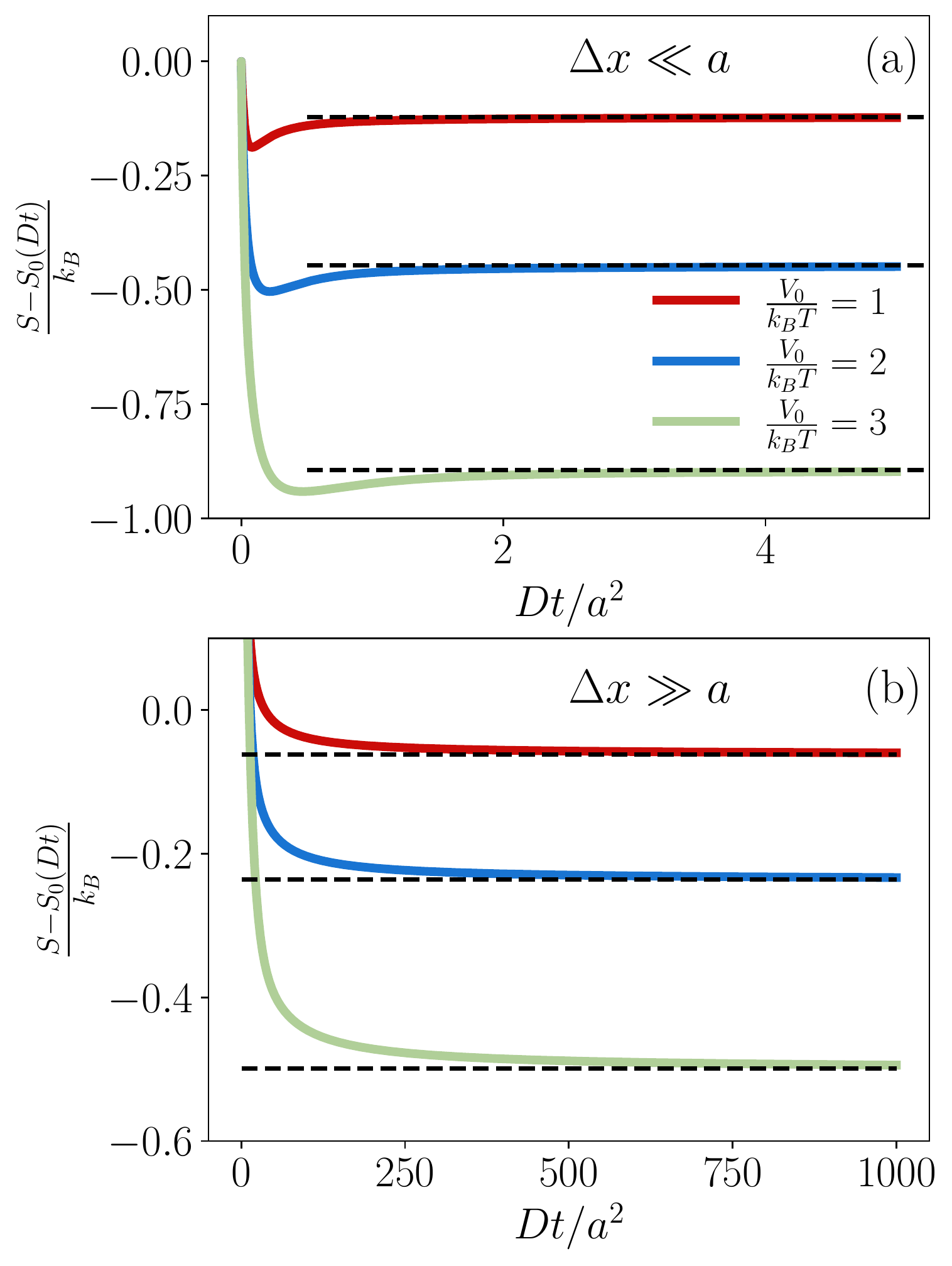}
		\caption{ \new{The entropy difference (solid lines) between that of a system with a cosine potential as Eq.\,(\ref{eq:cosine-pot-def}), $S$, and that of a free-particle with diffusion constant $D$, $S_0(D t)$, for very small and very large bin size: (a) $\Delta x = a/100$, and (b) $\Delta x = 10 a$.
        For long times, the entropy difference in the case of small bin size (panel (a)) becomes time-independent and is described by Eq.\,(\ref{eq:DeltaS_1}), which is very reminiscent of standard equilibrium within a unit cell.
        In the case of large bin size (panel (b)),  the entropy difference also becomes time-independent, as described by Eq.\,(\ref{eq:DeltaS_2}).
        The three values of $V_0/k_B T$ used are shown in the legend in panel (a).}  \label{fig:entropy}}
	\end{figure}

	We consider two isolated systems that started their motion at the same time with the same initial conditions, one with a periodic potential $V(x)$ (and energy $E$), and the other a free-particle $V(x) = 0$. All other parameters, temperature $T$ and bare diffusion constant $D$ are identical. We also assume we have acquired information about these systems using the same bin size $\Delta x$. For small bin size, $\Delta x \ll a$, The relative entropy is 
	\begin{eqnarray}
	    \frac{\Delta S}{k_B} =\frac{ S - S_0(D t)}{k_B} = \frac{E}{k_B T} \label{eq:DeltaS_1} \, ,
	\end{eqnarray}
	which only depends on the internal energy.
	In the limit of large bin size, $\Delta x \gg a$, the entropy difference becomes
	\begin{eqnarray}
	    \frac{\Delta S}{k_B} = \frac{S_0 (D^* t) - S_0(Dt)}{k_B} = \frac{1}{2} \ln \left[ \frac{D^*}{D} \right] \, . \label{eq:DeltaS_2}
	\end{eqnarray}
	As we would expect, the coarse-grained entropy does not depend on the internal energy, unlike Eq.\,(\ref{eq:DeltaS_1}). In Fig.\,(\ref{fig:entropy}) we plot the direct numerical calculation of Eq.\,(\ref{eq:S-general}) compared to our prediction of both small and large bin size limits in Eqs.\,(\ref{eq:DeltaS_1}) and (\ref{eq:DeltaS_2}).
	
	\subsection{Relative to the coarse-grained equivalent}

    As stated above, in Eq.\,(\ref{eq:S-theo-2}), we have a contribution equivalent to the entropy of a free particle with a renormalized diffusion constant, $S_0(D^* t)$. This time-dependent contribution is what we obtain by coarse-graining the system considering $\Delta x \gg a$, which leads to a Gaussian PDF with variance increasing as $\sqrt{2 D^* t}$, where we remark that $D^*$ can be measured using the mean square displacement (see Fig.\,\ref{fig:moments} and Eqs.\,(\ref{eq:msd}) and (\ref{eq:msd-time})).
    
    We can define another entropy difference, one that is the difference between the fine structure entropy in Eq.\,(\ref{eq:S-theo-2}) to the coarse-grained entropy in Eq.\,(\ref{eq:S-coarsed}), $S_0(D^*t)$, given by
	\begin{eqnarray}
	    \frac{\Delta S}{k_B} &=& \frac{S - S_0(D^* t)}{k_B} = \frac{E - F_a}{k_B T} \, \label{eq:DS-normal-vs-coarse} .
	\end{eqnarray}
    This is an extension of the regular entropy definition from the standard equilibrium statistical physics formula $S = \frac{E - F_a}{T}$. We also highlight that Eq.\,(\ref{eq:DS-normal-vs-coarse}) can be obtained from the results of a single system.
	
	\subsection{Different temperatures}

    Another possibility is to consider two systems that started at the same time but with different temperatures $T_1$ and $T_2$. Due to the difference in temperature, the systems will have different bare (without the periodic potential) diffusion constants $D_1$ and $D_2$ and two different internal energies $E_1$ and $E_2$. Their entropy difference, for long times, becomes
	\begin{eqnarray}
	    \frac{\Delta S}{k_B} &=& \frac{1}{2} \ln \left[ \frac{D_1}{D_2} \right] + \frac{E_1}{k_B T_1} - \frac{E_2}{k_B T_2} \, , \label{eq:DS-E1-T1-E2-T2}
	\end{eqnarray}  
	where we have a contribution from the macroscopic properties of the system, with the log of the ratio of diffusion constants and a microscopic contribution with the internal energies. This is due to the fact that both diffusion constant $D$ and internal $E$ depend on the temperature $T$.
	
	\subsection{Different internal energies}
	
	We now consider two systems with the same temperature $T$ (and the same bare diffusion constant $D$) but with different periodic potentials that lead to different internal energies $E_1$ and $E_2$. Using Eq.\,(\ref{eq:S-theo-1}), we obtain
    \begin{eqnarray}
        \frac{\Delta S}{k_B} &=&  \frac{E_1 - E_2}{k_B T} \, .
    \end{eqnarray}
	This expression, similar to Eq.\,(\ref{eq:DeltaS_1}), only depends on the temperature and the internal energies.
	
	\subsection{Different bin sizes}
	
	Lastly, we consider a single system and compare the entropy for large versus small bin size. In the limit $\Delta x_1 \ll a$, the entropy, which we label $S_1$, is given by Eq.\,(\ref{eq:S-theo-1}) and in the limit $\Delta x_2 \gg a$, the entropy, $S_2$ is given by Eq.\,(\ref{eq:S-coarsed}). The difference between these entropies is given by
	\begin{eqnarray}
	    \frac{\Delta S}{k_B} &=& \frac{S_1 - S_2}{k_B} = - \ln \frac{\Delta x_1}{\Delta x_2} + \frac{E - F_a}{k_B T} \, . \label{eq:DS-different-bins}
	\end{eqnarray}
	So far we managed to avoid entropies depending on the bin size by comparing systems where we acquire information using identical bin sizes. Since we have full knowledge of the bin size, we can use the expression in Eq.\,(\ref{eq:DS-different-bins}) to write a consistent definition of relative entropy for long times.
	It should be noted that the relationship between entropy production in out-of-equilibrium systems and its coarse-grained counterpart has already been extensively studied \cite{Esposito2012,Alonso-Serrano2017,Busiello2019,Chakraborti2022,Fiorelli2022}.
	
	\section{Eigenfunction derivation\label{sec:eigen}}
	
	We now present an eigenfunction derivation of the PDF in Eq.\,(\ref{eq:Pt-theoretical}). The probability density $P_t(x)$ can be written as an expansion of eigenfunctions as
	\begin{eqnarray}
		P_t(x) &=& \sum_{\{k\}} a_k \psi_k(x) e^{-\frac{V(x)}{2 k_B T}}e^{- \lambda_k D t} \, , \label{eq:eigenfunction-expansion}
	\end{eqnarray}
	where $\lambda_k$ is the eigenvalue associated with the eigenfunction $\psi_k(x)$ and $a_k$ sets the initial condition and ensures the normalization of $P_t(x)$. The eigenfunctions $\psi_k(x)$ are the solutions of the Schr\"odinger equation \cite{Risken1989}
	\begin{eqnarray}
		- \partial_x^2 \psi_k(x) + V_S(x) \psi_k(x) = \hat{H}_S \psi_k(x) = \lambda_k \psi_k(x) ~,~ ~ \label{eq:schrd}
	\end{eqnarray}
	where we have defined the effective Hamiltonian operator $\hat{H}$ and the effective potential as
	\begin{eqnarray}
		V_S(x) &=& \frac{V'(x)^2}{4 k_B^2 T^2} - \frac{V''(x)}{2k_B T} \, .
	\end{eqnarray}
	We remark that $V_S(x)$ is also $a$-periodic and that the eigenvalues $\lambda_k$ are the energy levels of Eq.\,(\ref{eq:schrd}). From Eq.\,(\ref{eq:eigenfunction-expansion}) it becomes clear that large values of $\lambda_k$ are going to have their contributions to $P_t(x)$ suppressed by the $e^{- \lambda_k D t}$ term. At long times, only the smaller values of $\lambda_k$ are going to contribute. 
	
	Using Bloch's theorem, we find the solutions $\psi_k(x)$ of the Schr\"odinger Equation, Eq. \,(\ref{eq:schrd}), and using Eq.\,(\ref{eq:eigenfunction-expansion}), we obtain
	\begin{eqnarray}
	    P_t(x) & \approx & \frac{e^{-\frac{V(x)}{k_B T}} e^{- \frac{x^2}{4 D^* t}}}{\sqrt{4 \pi D t}} \left[ 1 - \frac{x \, U_1(x) + U_2(x)}{2 D^* t} \right], \label{eq:Pt_final_first}
	\end{eqnarray}
	which is the same expression as in Eq. (\ref{eq:Pt-theoretical}). \new{The details of the calculations can be found in Appendix \ref{app:eigenfunctions}.}
    We remark that this solution is valid for any periodic potential, the unit cell of the potential does not need to be symmetric, and we may treat the problem of particles starting at $x=x_0$ by translating the potential and using the same expressions.

	\section{Final Remarks \label{sec:final-remarks}}

        We have studied herein the properties of \new{four classes (i-iv) of} observables for overdamped Brownian particles in a periodic potential. In the long-time limit, this system approaches a Boltzmann-Gibbs steady state, as described by Eq.\,(\ref{eq:infinite-density-def}). The key feature is that the Boltzmann-Gibbs factor, $e^{-V(x)/k_B T}$,  is non-normalizable. This implies unusual ergodic and thermodynamic properties of the system.
        
        Despite the nonbinding nature of the periodic potential and the absence of a true normalized Boltzmann-Gibbs equilibrium solution for long times, observables that have the same periodicity as the potential \new{(i)} will have their ensemble averages converge to the expected values for a system in equilibrium in a single unit cell of the periodic structure and will follow regular equilibrium ergodicity. An example of such an observable is the internal energy, see Eq.\,(\ref{eq:internal-energy}) and Eq.\,(\ref{eq:prediction-Ei}). On the other hand, there is a different class of integrable observables \new{(ii)}, such as the indicator function in Eq.\,(\ref{eq:indicator-func}), whose averages do not follow equilibrium or ergodicity in its usual sense, but can still be calculated, as we can see from Eq.\,(\ref{eq:infinite-avg}), using the Boltzmann-Gibbs factor. Unlike regular equilibrium, the ensemble average of the time average of these observables shows a doubling effect, as shown in Eq.\,(\ref{eq:relation-between-averages}). The distribution of the time averages do not become narrower with increasing measurement time, as we showed in Fig. \ref{fig:observables-ergodic}(b). In this case, we demonstrated how the Darling-Kac theorem yields the statistics of time averages. Still, the key issue is that these are evaluated with the Boltzmann-Gibbs factor. \new{The virial observable (iii) has some unique properties as it marks a transition in the ergodic properties of observables, as demonstrated in Fig. \ref{fig:variances}.}
        A different class of observables are \new{(iv)} the positional moments. These are insensitive to the fine scale, and they exhibit standard ergodicity in the mean square displacement sense.
        
        Given that the system, according to Eq.\,(\ref{eq:infinite-density-def}), reaches a non-normalized Boltzmann-Gibbs state, we proceeded to unravel some of the thermodynamical relations in this model. We showed in Eq.\,(\ref{eq:virial-theo}) that the virial theorem is controlled by the ratio $D^*/D$, which clearly depends on the dynamical behavior of the system, namely, on the mean square displacement.
        The entropy difference between an ensemble of non-interacting particles in the periodic field, and an ensemble of freely diffusing particles, or two systems with different temperatures is also related to the ratio between diffusion constants, as shown in Eq.\,(\ref{eq:DeltaS_2}) for large $\Delta x \gg a$, where $\Delta x$ is the bin size. For small $\Delta x \ll a$, the entropy difference between the entropy of a system and the coarse-grained equivalent from the same system shows how $F_a=T \Delta S - E$ (see Eq.\,(\ref{eq:DS-normal-vs-coarse})) where $F_a$ is the free energy of a particle in one lattice unit. This free energy is equivalently obtained from $Z_a$, the partition function defined in a single unit cell.
        This relation between entropy, average energy, and free energy, is very much reminiscent of the basic relation between these thermodynamic functions as found in ordinary statistical physics.
        The fundamental difference is that the entropy of both systems is always increasing with time since the systems under study are unbounded, while for finite systems the entropy will eventually saturate to a fixed value.
        
        Using an eigenfunction expansion, we have extended the Sivan-Farago expression of the PDF of particles in general periodic potentials, Eq.\,(\ref{eq:Pt-theoretical}). This is an accurate description in the long time ($t \gg a^2/D^*$) of the spreading packet of particles. With the PDF obtained, it is possible to investigate both the macroscopic behavior, that is, the effective diffusion constant $D^*$, as well as the microscopic intra-well behavior of the particles.
        
       Important desirable extensions to this work include the study of the solutions for systems in higher dimensions as well as for underdamped motion in a periodic potential. Another possible direction is to investigate the effects of many interacting particles, such as single file diffusion systems \cite{Taloni2006} or systems with periodic forces under non-thermal noises \cite{Hanggi2020}. The machinery of stochastic thermodynamics has not been studied here and can be expected to yield further insights.
        
        \begin{acknowledgments} The support of Israel Science Foundation’s Grant No.
        1614/21 is acknowledged.
    \end{acknowledgments}

        \appendix 

        \new{
        \section*{Appendix: Eigenfunction derivation \label{app:eigenfunctions}}
        
    We will now show the detailed derivation of the PDF using the eigenfunction expansion in Eq.\,(\ref{eq:eigenfunction-expansion}). According to Bloch's theorem, the solutions of a Schr\"odinger Equation with a periodic potential, such as Eq. \,(\ref{eq:schrd}), can be written as
	\begin{eqnarray}
		\psi_k(x) = e^{i k x} u_k(x) \label{eq:bloch-theo} \, ,
	\end{eqnarray}
	where $k$ is the wavevector of the entire lattice and $u_k(x)$ is an $a$-periodic function. We imagine our system in a box of size $2L$ (boundaries at $x = \pm L$), which leads to a discrete eigenspectrum of the operator $\hat{H}_S$ and $k_n = {n \pi}/{L}$. The eigenfunctions are orthogonal, that is, $\int_{-L}^{L} \psi_k^*(x)\psi_{k'}(x) dx \propto \delta_{k\,k'}$, where $\delta_{k\,k'}$ is Kronecker's delta. We can use the initial probability $P_0(x) = \delta(x)$, together with the orthogonality of the eigenfunctions, to write that
	\begin{eqnarray}
	    a_k &=& \frac{\int_{-L}^{L} \psi_k^*(x) P_0(x) e^{\frac{V(x)}{2 k_B T}} dx}{\int_{-L}^{L} \psi_k^*(x) \psi_k(x) dx} \nonumber = \frac{e^{\frac{V(0)}{2k_B T}} \psi_k^*(0)}{\int_{-L}^{L} \psi_k^*(x) \psi_k(x) dx} \, . \\
	\end{eqnarray}
	
	In order to solve the Schr\"odinger equation, which is a second-order linear equation, we must obtain two independent solutions, $\varphi(x)$ and $\phi(x)$. To ensure that these solutions are linearly independent, it is sufficient to have boundary conditions $\varphi(0) = 1$, $\varphi'(0) = 0$ and $\phi(0) = 1$, $\phi'(0) = 1/a$, as it ensures that $\varphi(x)$ and $\phi(x)$ are not proportional to one another and are not null. As we are only interested in the long-time limit, for which only small $\lambda_k$ contribute, these solutions will also be expressed as a series expansion of $\lambda_k$, that is, 
	\begin{eqnarray}
	    \varphi(x) &\approx & \varphi_0(x) + \lambda_k \varphi_1(x) \label{eq:exp-varphi} \\
	    \phi(x) &\approx & \phi_0(x) + \lambda_k \phi_1(x) \, . \label{eq:exp-phi}
	\end{eqnarray}
	This approximation is only valid for values of $x$ where $\varphi_0(x) \gg \lambda_k \varphi_1(x)$, which, as we will see later in this section, is equivalent to the restriction $kx \ll 1$. Because of Bloch's theorem, we only require the solutions to be valid in the range $0 \leq x \leq a$, where such restriction is easily satisfied. We remark that the series expansion in Eqs.\,(\ref{eq:exp-varphi}) and\,(\ref{eq:exp-phi}) must follow the boundary condition up to $O(\lambda_k)$.
	
	The zeroth ($\lambda_k = 0$) order solutions of Eq.\,(\ref{eq:schrd}) are
	\begin{eqnarray}
	    \varphi_0(x) &=& e^{- \frac{V(x)}{2k_B T} + \frac{V(0)}{2k_B T}} \label{eq:varphi-0} \, \\
	    \phi_0(x) &=& \frac{e^{- \frac{V(x)}{2k_B T} - \frac{V(0)}{2k_B T}}}{a} \int_0^x e^{\frac{V(y)}{k_B T}} dy \, , \label{eq:phi-0}
	\end{eqnarray}
	as they satisfy $\hat{H}_S \varphi_0(x) = 0$ and $\hat{H}_S \phi_0(x) = 0$. We chose the origin to have null derivative $V'(0) = 0$, this ensures that the boundary conditions $\varphi_0(0)= a \phi'(0) = 1$, $\varphi_0'(0) = \phi_0(0) = 0$ are satisfied. We remark that this last restriction is actually unnecessary and the final result is general. The first order solutions are obtained by plugging Equations\,(\ref{eq:varphi-0}) and (\ref{eq:phi-0}) in Eq.\,(\ref{eq:schrd}), that is, $\hat{H}_S \varphi_1(x) = -\lambda_k \varphi_0(x)$ and $\hat{H}_S \phi_1(x) = -\lambda_k \phi_0(x)$. These solutions are
	\begin{eqnarray}
	    \varphi_1(x) &=& - e^{-\frac{V(x)}{2 k_B T} +\frac{V(0)}{2 k_B T} }  \nonumber \\
        &~& \int_0^x \int_0^{y_1} e^{\frac{V(y_1)}{k_B T}-\frac{V(y_2)}{k_B T}} dy_2 dy_1 \label{eq:varphi-1} \\
	    \phi_1(x) &=& - \frac{e^{-\frac{V(x)}{2 k_B T} - \frac{V(0)}{2 k_B T} }}{a}\label{eq:phi-1} \\\nonumber 
	    &~& \int_0^x \int_0^{y_1} \int_0^{y_2} e^{\frac{V(y_1)}{k_B T} -\frac{V(y_2)}{k_B T}  + \frac{V(y_3)}{k_B T}}   d y_3 dy_2 dy_1 . 
	\end{eqnarray}
	We have that $\varphi_1(0) = \varphi_1'(0) = \phi_1(0) = \phi_1'(0) = 0$, and therefore, the boundary conditions of $\varphi(x)$ and $\phi(x)$ are satisfied up to $O(\lambda_k)$. 
	
	\subsection{The eigenvalue spectrum $\lambda_k$}
	
	The eigenvalues can be obtained using the symmetry operator $\hat{T}_a$, the translation by a length $a$, that is, $\hat{T}_a \psi_k(x) = \psi_k(x+a) = e^{ika}\psi_k(x)$. Clearly, the eigenvalues of $\hat{T}_a$ are $\nu_\pm = e^{\pm i k a}$, and $\psi_k(x)$, as defined in Eq.\,(\ref{eq:bloch-theo}), are the eigenfunctions. In the basis of $(\varphi(x),\phi(x))$, the translation can be described as a linear combination, and therefore, it is possible to write the matrix representation of the operator as
	\begin{eqnarray}
	    \hat{T}_a = \begin{pmatrix}
	    \varphi(a) & \phi(a) \\
	    a \varphi'(a) & a \phi'(a) 
	    \end{pmatrix} \, , \label{eq:matrix-Ta}
	\end{eqnarray}
	which we use to obtain the eigenvalues $\nu$ from
	\begin{eqnarray}
	   \nu^2 - (\varphi(a) + a \phi'(a)) \nu + 1 &=& 0 \, .
	\end{eqnarray}
	From our series expansion for $\varphi(x)$, in Eq.\,(\ref{eq:exp-varphi}), and $\phi(x)$, in Eq.\,(\ref{eq:exp-phi}), we obtain that,
	\begin{eqnarray}
	    \varphi(a) &=& 1 - {\lambda_k} \int_0^{a} e^{\frac{V(y_1)}{k_B T}} \int_0^{y_1} e^{-\frac{V(y_2)}{k_B T}} dy_2 dy_1 \\
	    \phi'(a) &=&  \frac{1}{a} - \frac{\lambda_k}{a} \int_0^{a} e^{\frac{V(y_1)}{k_B T}} \int_{y_1}^a e^{-\frac{V(y_z)}{k_B T}} dy_2 dy_1 \, ,
	\end{eqnarray} 
	and the sum $\varphi(a) + a \phi'(a) = 2 - \lambda_k \left\langle e^{{V}/{k_B T}} \right\rangle_a$.
	We obtain the series for the eigenvalues as
    \begin{eqnarray}
	    \nu_{\pm} & \approx & 1 \pm i a \sqrt{\lambda_k} \left\langle e^{\frac{V}{k_B T}} \right\rangle_a - \frac{a^2 \lambda_k}{2} \left\langle e^{\frac{V}{k_B T}} \right\rangle_a^2 \, . \label{eq:eigenvalues-lambda}
	\end{eqnarray}
	We match this solution with the eigenvalues of $\hat{T}_a$, $\nu_\pm = e^{\pm i k a} \approx 1 \pm i a k - {k^2 a^2}/{2}$, to obtain the eigenvalues $\lambda_k$ as
	\begin{eqnarray}
	    \lambda_k &\approx & \frac{k^2}{\left\langle e^{\frac{V}{k_B T}} \right\rangle_a^2} = \frac{D^*}{D} k^2 \, . \label{eq:energy-band}
	\end{eqnarray}
	This is the expected result that describes a free particle, with a renormalized diffusion constant.
	
	\subsection{The eigenfuncions}
	
	From the matrix representation of $\hat{T}_a$ in Eq. (\ref{eq:matrix-Ta}), we can immediately conclude that the eigenfunctions must be
	\begin{eqnarray}
	    \psi_{\pm k} & = & \varphi(x) + \frac{\nu_\pm - \varphi(a)}{\phi(a)} \phi(x) \, ,
	\end{eqnarray}
	where we replace $\nu_\pm$ obtained in Eq. (\ref{eq:eigenvalues-lambda}) to write
	\begin{eqnarray}
	    \frac{\nu_\pm - \varphi(a)}{\phi(a)} &=& \pm \frac{i k a}{\phi_0(a)} - \frac{k^2}{\phi_0(a)} \left[ \frac{a^2}{2} - \frac{D^*}{D} \varphi_1(a) \right] \nonumber \\
	    &=& \pm i k a \, \sqrt{\frac{D^*}{D}} \, e^{\frac{V(0)}{k_B T}} - e^{\frac{V(0)}{k_B T}} \frac{k^2 a  D^*}{D} C_0 \nonumber \\ \, ,
	\end{eqnarray}
	where $C_0$ is the same as we obtained in Eq.\,(\ref{eq:C0}). We now write
	\begin{eqnarray}
	    \psi_{\pm k}(x) &=& \varphi_0(x) \pm i k \sqrt{\frac{D^*}{D}} e^{\frac{V(0)}{k_B T}}  \phi_0(x) + \frac{D^* k^2}{D} \varphi_1(x) \nonumber \\
	    &~& - e^{\frac{V(0)}{k_B T}} \frac{k^2 a  D^*}{D} C_0 \phi_0(x) \, . \label{eq:psi-pm-phis}
	\end{eqnarray}
	We remark that these eigenfunctions do not need to be normalized, as the $a_k$ terms in Eq. (\ref{eq:eigenfunction-expansion}) will ensure normalization. In the limit of small $k$, we can write the periodic functions $u_k(x)$ of the Bloch waves (see Eq. (\ref{eq:bloch-theo})) as a series of $k$, that is, $u_k(x) = u_0(x) + i k u_1(x) - k^2 u_2(x) + O(k^3)$. From Eq.\,(\ref{eq:bloch-theo}), using Eq.\,(\ref{eq:psi-pm-phis}), we can write
    \begin{eqnarray}
        u_k(x) &=& e^{- i k x} \psi_k(x) \nonumber \\
        & \approx & \varphi_0(x) + ik \left( a \sqrt{\frac{D^*}{D}} e^{\frac{V(0)}{k_B T}} \phi_0(x) - \varphi_0(x) x \right) \nonumber + \\
        &~& +k^2(...) 
        \, . \label{eq:bloch-expected}
    \end{eqnarray}
    For simplicity, we have omitted the $u_2(x)$ term. By matching the same order of $k$ on the left and right-hand sides of Eq. (\ref{eq:bloch-expected}), we obtain
	\begin{eqnarray}
	    u_0(x) &=& \varphi_0(x) \label{eq:varphi0} \,  \\
        u_1(x) &=& u_0(x) \left[ a  \frac{\int_0^x e^{\frac{V(y)}{k_B T}} dy}{\int_0^a e^{\frac{V(y)}{k_B T}} dy} - x \right] = u_0(x) U_1(x), \nonumber \\ \label{eq:phi0} 
	\end{eqnarray}
    where we can see that $U_1(x)$ is the same function as the one obtained in Eq.\,(\ref{eq:U1-def}). The expression for $u_2(x)$, which we have omitted for simplicity, can also be derived from Eq. (\ref{eq:psi-pm-phis}).

    \subsection{Obtaining the PDF}
	
	The last ingredient to complete the eigenfunction expansion is the normalization $a_k$, which we can obtain through \cite{Risken1989}
	\begin{eqnarray}
	    a_k &=& \frac{e^{\frac{V(0)}{2k_B T}} \psi_k^*(0)}{\int_{-L}^{L} |u_k(x)|^2 dx} \approx \frac{e^{\frac{V(0)}{2k_B T}} \psi_k^*(0)}{2L \left\langle |u_k|^2 \right\rangle_a} \nonumber \\ 
	    & \approx & \frac{e^{-\frac{V(0)}{2k_B T}} \psi_k^*(0)}{2L} \sqrt{\frac{D^*}{D}}\left( 1 - k^2 \frac{\left\langle u_1^2 + u_0 u_2 \right\rangle_a}{\left\langle u_0^2 \right\rangle_a} \right) , \nonumber \\
	\end{eqnarray}
	where we identify the normalization constant found in Section \ref{sec:asymptotic-solution}, $C_1 = {\left\langle u_1^2 + u_0 u_2 \right\rangle_a}/{\left\langle u_0^2 \right\rangle_a}$.
    In the limit of $L \to \infty$, we can replace the sum in $k$ modes by an integral as $ \sum_{k} = \int \frac{L dk}{\pi}$, and Eq. (\ref{eq:eigenfunction-expansion}) becomes
	\begin{eqnarray}
		P_t(x) & \approx & e^{-\frac{V(x)}{2 k_B T}} \sqrt{\frac{D^*}{D}}  \int_{-\infty}^{\infty} \frac{dk}{2 \pi}  e^{- k^2 D^* t} e^{i kx} \nonumber \\
		&~& \bigg\{ u_0(x) + i k u_1(x)  - k^2 [u_2(x) + u_0(x) C_1]) \bigg\} \nonumber \\
		& \approx & e^{-\frac{V(x)}{k_B T}} \sqrt{\frac{D^*}{D}}  \int_{-\infty}^{\infty} \frac{dk}{2 \pi}  e^{- k^2 D^* t} e^{i kx} \nonumber \\
		&~&  \bigg\{ 1 + i k U_1(x)  - k^2 U_2(x)) \bigg\} \, , 
	\end{eqnarray}
	 where we simplified the expression defining $U_2(x) = u_2(x)/u_0(x) + C_1$, which is identical to the one in Eq.\,(\ref{eq:def-tau}). Performing the integral, we obtain
	\begin{eqnarray}
	    P_t(x) & \approx & \frac{e^{-\frac{V(x)}{k_B T}} e^{- \frac{x^2}{4 D^* t}}}{\sqrt{4 \pi D t}} \left[ 1 - \frac{x \, U_1(x) + U_2(x)}{2 D^* t} \right] \, , \label{eq:Pt_final_first}
	\end{eqnarray}
	where we reach the same expression as in Eq. (\ref{eq:Pt-theoretical}).
    
    We remark that when $V'(0) \neq 0$, it is necessary to add terms in the expression for $\varphi(x)$ to ensure that $\varphi'(0) = 0$. These extra terms do not affect the result of the energy band in Eq. (\ref{eq:energy-band}) or the expression for the periodic functions in Equations (\ref{eq:varphi0}) and (\ref{eq:phi0}), as they will always be canceled out. }

\end{document}